\documentclass{arXiv}

\usepackage{physics}
\usetikzlibrary{quantikz}
\usepackage{textcomp}
\usepackage{amsthm}
\usepackage{todonotes}

\newcommand{\R}{\mathbb{R}}
\newcommand{\C}{\mathbb{C}}
\newcommand{\T}{\mathbf{T}}
\newcommand{\U}{\mathbf{U}}
\newcommand{\G}{\mathbf{G}}
\newcommand{\PROB}{\mathbf{P}}
\newcommand{\CNOT}{\mathrm{CNOT}}
\newcommand{\CCNOT}{\mathrm{CCNOT}}
\newcommand{\boldH}{\mathbf{H}}
\newcommand{\PSI}{\mathbf{\Psi}}

\newcommand{\diag}{\mathrm{diag}}
\newcommand{\core}[1]{\left\llbracket\, \begin{matrix} #1 \end{matrix} \,\right\rrbracket}

\usepackage{xcolor}
\usepackage{verbatim}

\usepackage[sorting=none, style=numeric, firstinits=true, maxbibnames=15, citestyle=numeric-comp]{biblatex}
\bibliography{references}

\begin{document}

\title{\huge Low-rank tensor decompositions of quantum circuits}

\abstract{%
Quantum computing is arguably one of the most revolutionary and disruptive technologies of this century.
Due to the ever-increasing number of potential applications as well as the continuing rise in complexity, the development, simulation, optimization, and physical realization of quantum circuits is of utmost importance for designing novel algorithms.
We show how matrix product states (MPSs) and matrix product operators (MPOs) can be used to express certain quantum states, quantum gates, and entire quantum circuits as low-rank tensors.
This enables the analysis and simulation of complex quantum circuits on classical computers and to gain insight into the underlying structure of the system.
We present different examples to demonstrate the advantages of MPO formulations and show that they are more efficient than conventional techniques if the bond dimensions of the wave function representation can be kept small throughout the simulation.
}

\keywords{quantum computing, quantum simulation, matrix product states, matrix product operators, tensor trains}

\msc{
15A69, 
47N50, 
65C05, 
68W40, 
81P68  
}
\doi{}
\author{Patrick Gel\ss}{AI in Society, Science, and \\Technology, Zuse Institute Berlin, \\ Berlin 14195, Germany \\[0.1cm] Institute of Mathematics,\\Freie Universität Berlin,\\Berlin 14195, Germany}
\author{Stefan Klus}{School of Mathematical \& \\Computer Sciences,\\Heriot-Watt University,\\Edinburgh EH14 4AS, UK}
\author{Sebastian Knebel}{AI in Society, Science, and \\Technology, Zuse Institute Berlin, \\ Berlin 14195, Germany}
\author{Zarin Shakibaei}{AI in Society, Science, and \\Technology, Zuse Institute Berlin, \\ Berlin 14195, Germany}
\author{Sebastian Pokutta}{AI in Society, Science, and \\Technology, Zuse Institute Berlin, \\ Berlin 14195, Germany \\[0.1cm] Institute of Mathematics, \\ Technische Universität Berlin, \\ Berlin 10623, Germany}
\email{p.gelss@fu-berlin.de}

\maketitle

\section{Introduction}

In the early 1980s, Benioff~\cite{Benioff1980}, Manin~\cite{Manin1980}, and Feynman~\cite{Feynman1982} proposed a novel computing paradigm based on quantum mechanics, which later became known as \emph{quantum computing}.
In classical computing, the smallest unit of information is the binary digit or bit representing a logical state with one of two possible values. In most quantum computers today, the smallest unit of information is the \emph{quantum bit} or \emph{qubit}, whose state can be described by a linear combination of two distinct states according to quantum mechanics' principle of superposition. Multiple qubits can be combined to quantum registers, in which the involved qubits can further interact with each other by exploiting the principle of entanglement.
Quantum circuits as models of computation on quantum computers are built up as combinations of \emph{quantum (logic) gates}, which are fundamental quantum circuits operating on single qubits or quantum registers.
The operations are reversible and described by unitary operators.

The interest in quantum computing has increased significantly over the last years, both in technical realizations and in the theory of quantum algorithms.
Different hardware architectures have been considered~\cite{Monroe2014, Saffman2016, Gaita2019, Zhong2020}.
Today, quantum computers are a reality, where in particular the superconducting type~\cite{Clarke2008} has been built by companies such as IBM, Google, and Intel.
With increasing frequency, they are used for such versatile applications as machine learning~\cite{Biamonte2017b, Zhang2020}, cryptography~\cite{Bernstein2017, Pirandola2020}, chemistry~\cite{Kassal2008, McArdle2020}, and finance~\cite{Rebentrost2018, Coyle2021}.
One of the objectives of quantum computing is to outperform classical computers on problems that are intractable for them.
This is known as \emph{quantum supremacy}, a point which Google~\cite{Arute2019} claims to have achieved for the first time in 2019 and which has generated extensive scientific and public attention.
However, so far only few quantum algorithms have been developed that are able to outperform known classical counterparts on real-world problems. Important examples of such algorithms are \emph{Shor's algorithm} for integer factorization~\cite{Shor1994}, \emph{Grover's algorithm} for searching unsorted databases~\cite{Grover1996}, and \emph{Simon's algorithm} for finding periods of functions~\cite{Simon1997}.
The former two algorithms exhibit a (super-) polynomial speed-up and the latter an exponential speed-up compared to the best known classical algorithms.

The development of quantum algorithms requires concepts such as genetic programming~\cite{Spector2004} or quantum walks~\cite{Kendon2006} that are able to express the logic of complex quantum circuits and translate classical mathematical operations into instructions that can be executed on quantum computers and vice versa.
For this, the simulation of quantum computers on classical hardware is of utmost importance.
Although this so-called \emph{quantum simulation} helps to explore the capabilities of quantum computers and to understand their computational benefits as well as limitations, it typically leads to severe numerical problems.
This is mainly due to the fact that the storage consumption and the computational costs grow exponentially with the number of qubits, often referred to as the \emph{curse of dimensionality}.
Improved methods are needed to overcome this problem.

One approach to tackle high-dimensional quantum systems is to represent them as tensor networks.
Various methods for enhancing quantum simulations based on \emph{matrix product states} (MPS)~\cite{Banuls2006, Lanyon2017, Ganahl2017, DeNicola2021, Zhou2020}, \emph{projected entangled pair states} (PEPS)~\cite{Guo2019}, and the \emph{multi-scale entanglement renormalization ansatz} (MERA)~\cite{Luchnikov2021} have been proposed. 
MPSs were already introduced in quantum physics in 1987~\cite{Affleck1987} and were reinvented in applied mathematics around 2009 under the name \emph{tensor trains} (TTs)~\cite{Oseledets2009a, Oseledets2009b}. Besides their application in quantum mechanics, e.g.,~\cite{Veit2017, Gelss2022}, TTs have become a widely studied concept. By now, they are also applied in other scientific fields such as dynamical systems~\cite{Klus2018, Goessmann2020} and quantum machine learning~\cite{Huggins2019}.
Different efficient algorithms for TTs have been developed, ranging from least-squares solvers to data-driven methods like ALS~\cite{Holtz2012}, SALSA~\cite{Grasedyck2019}, ARR~\cite{Klus2019}, MANDy~\cite{Gelss2019}, AMUSEt~\cite{Nueske2021}, and tgEDMD~\cite{Luecke2021}.
As shown below, the tensor train format allows for compact expressions, systematic decompositions, and direct analysis of the entanglement of quantum states.

Until now, tensor decompositions have been mainly used to represent quantum states. A natural extension of MPSs are \emph{matrix product operators} (MPOs).
MPOs have also recently been used to represent quantum systems, but only for one- and two-qubit operators either satisfying the nearest-neighbor constraint~\cite{Almudever2017}, without focus on closed expressions~\cite{Ran2020, Torlai2020}, or for representing specific sequences of bipartite unitary operations~\cite{Saberi2011, Saberi2013}.
To our knowledge, extensive studies of explicit MPO representations of complex quantum circuits have not been considered yet.

The aim of this work is to represent quantum states as MPS as well as quantum gates and entire quantum circuits as MPO decompositions.
Specifically, it targets the derivation of low-rank MPO representations of (networks of) quantum gates acting on arbitrary qubits within the quantum register, similar to the MPS representation of quantum states.
Besides demonstrating the use of the MPS/MPO formalism for gaining insights into the network structure of quantum systems, also new avenues for tensor-based research might be opened up, in particular pertaining to the development and simulation of quantum algorithms by describing circuits as closed-form operators.

Several quantum circuits are examined by contracting MPS and MPO networks. As a result, the probability distribution for the measurement outcomes can be generated in two ways: either constructed directly as a tensor or efficiently sampled by using a generative sampling strategy~\cite{Ferris2012, Han2018}.
Furthermore, it is shown that various circuits can be expressed as weakly entangled MPOs.
In contrast to existing state-of-the-art simulation techniques, it is demonstrated that for some cases the runtime of quantum algorithms built on MPOs increases only linearly with the size of the register.
The specific contributions of the present study are as follows:
\begin{itemize}
 \item First, general MPO decompositions of single-qubit and (multi-)controlled quantum gates acting on $n$-qubit systems are derived without requiring the nearest-neighbor constraint, allowing interactions between arbitrary qubits in the quantum register.
 \item Second, we consider compact MPO representations of several fundamental quantum circuits as concatenations of multiple quantum gates.
 These MPO representations allow for the direct analysis of their entanglement structure and their action on quantum states.
 \item Finally, the capabilities of the proposed method are demonstrated by simulating different quantum algorithms and comparing the obtained results to results computed using Qiskit.
\end{itemize}
This paper is structured as follows: Section~\ref{sec: Preliminaries} recalls quantum states and quantum gates and briefly introduces the MPS/MPO/TT formats.
Section~\ref{sec: Quantum gates and circuits in TT formalism} demonstrates the application of MPOs to quantum system representations using the example of three well-known quantum algorithms.
Section~\ref{sec: Numerical experiments} provides numerical examples of MPS/MPO-based quantum simulations.
Section~\ref{sec: Conclusion and outlook} concludes with a summary of the main results and their potential impact on further research.
Appendix~\ref{app: Appendix} contains supplementary information and detailed derivations of MPO representations.

\section{Preliminaries}
\label{sec: Preliminaries}

Let $n\in\mathbb{N}$ and $D:=[d_1, \dots, d_n]^\top\in\mathbb{N}^n$ and consider the vector spaces $\mathbb{C}^{d_1},\dots,\mathbb{C}^{d_n}$.
Then, the tensor product (space) $\mathcal{V}:=\bigotimes_{i=1}^n\mathbb{C}^{d_i}$ is a complex vector space equipped with the $n$-multilinear map $\otimes:\bigtimes_{i=1}^n\mathbb{C}^{d_i}\to\mathcal{V}$, such that all bilinear maps out of $\bigtimes_{i=1}^n\mathbb{C}^{d_i}$ factor uniquely through $\otimes$.
If for all $i\in\{1,\dots,n\}$ the spaces $\mathbb{C}^{d_i}$ are equipped with bases $\{e^{(i)}_{x_i} \ \big|\ 1\leq x_i\leq d_i \}$, then $\{\bigotimes_{i=1}^n e^{(i)}_{x_i} \ \big|\ 1\leq x_1\leq d_1,\dots ,1\leq x_n\leq d_n \}$ is a basis of $\mathcal{V}$.
An element $T\in\mathcal{V}$ is called tensor and can be written as 
\begin{equation*}
    T = \sum_{x_1=1}^{d_1}\cdots\sum_{x_n=1}^{d_n} T_{x_1,\dots,x_n} \bigotimes_{i=1}^n e^{(i)}_{x_i},
\end{equation*}
where $T_{x_1,\dots,x_n}\in\mathbb{C}$. This representation is called \emph{canonical format}. For given bases, a tensor in the canonical format is uniquely determined by the multidimensional array 
\begin{equation*}
    \T:=[T_{x_1,\dots,x_n}]_{1\leq x_1\leq d_1, \dots, 1\leq x_n\leq d_n}\in\mathbb{C}^{d_1\times\cdots\times d_n}.    
\end{equation*}
In this work, with a slight abuse of language, this multidimensional array itself is called tensor of order $n$ with mode set $D$ and will be denoted by bold letters.
An element of a tensor $\T$ (as array) is addressed by subscript indices.
We both define $\C^D := \C^{d_1 \times \dots \times d_n}$ and the tensor product of an order-$n$ tensor $\T$ with an order-$m$ tensor $\U$ as
\begin{equation*}
    (\T \otimes \U)_{x_1, \dots, x_n, y_1, \dots , y_m} := \T_{x_1, \dots, x_n} \U_{y_1, \dots, y_m}
\end{equation*}
for any possible combination of mode indices.

\subsection{Quantum states}
\label{sec: Quantum states}

As a foundation for describing quantum circuits in MPS/MPO format, basic representations of quantum registers are recalled. To this end, define the Hilbert space $\mathcal{H}$ as $\mathbb{C}^2$ equipped with inner product to describe the state of a single qubit as a vector $\ket{x}\in\mathcal{H}$. A qubit has two computational basis states, i.e., $x\in\{0,1\}$, so that $\ket{0}=[1,0]^\top$ and $\ket{1}=[0,1]^\top$. A quantum register consisting of $n$ qubits is described as a vector $\ket{x}\in\mathcal{H}^{\otimes n}:=\bigotimes_{i=1}^n \mathcal{H}={(\mathbb{C}^2)}^{\otimes n}$. It is often written as
\begin{equation*}
 \ket{x} = \ket{x_1 x_2 \cdots x_n} := \ket{x_1} \otimes \dots \otimes \ket{x_n},
\end{equation*}
with $x_i \in \{0,1\}$ for all $i\in\{1,\dots ,n\}$.
By convention, the basis state index $x$ is the decimal number associated to the binary number that is encoded by the bit string $x_1x_2\cdots x_n$, with $x_{1}$ as its most significant bit.
A pure quantum state of $n$ qubits can then be written as
\begin{equation}\label{eq: quantum state}
    \ket{\psi} = \sum_{\mathclap{x_1, \dots , x_n = 0}}^1 \PSI_{x_1, \dots, x_n} (\ket{x_1} \otimes \dots \otimes \ket{x_n}), 
\end{equation}
where the tensor $\PSI \in \mathbb{C}^{(2^{\times n})}$ represents the wave function. That means, it contains the probability amplitudes corresponding to each basis state such that $\sum_{x_1,\dots, x_n=0}^{1} \abs{\PSI_{x_1, \dots, x_n}}^2 = 1$. 
A quantum state is called \emph{separable} if it can be written as a \emph{rank-one} tensor, i.e., a tensor product of $n$ components:
\begin{equation*}
 \ket{\psi} = \ket{\psi_1 \psi_2 \dots \psi_n} = \ket{\psi_1} \otimes \dots \otimes \ket{\psi_n} = \begin{bmatrix} \psi_{1,0} \\ \psi_{1,1} \end{bmatrix} \otimes \dots \otimes \begin{bmatrix} \psi_{n,0} \\ \psi_{n,1} \end{bmatrix},
\end{equation*}
where $\ket{\psi_i} = [\psi_{i,0}, \psi_{i,1}]^\top$ is a pure state of the $i$th subsystem. In this case, the entries of $\PSI$ in \eqref{eq: quantum state} are given by 
\begin{equation*}
    \PSI_{x_1, \dots, x_n} = \prod_{i=1}^n \psi_{i,x_i}. 
\end{equation*}
However, tensor products will be applied not only for expressing separable quantum states, but also highly \emph{entangled} ones. For this purpose, i.e., representing any given quantum state as a tensor network, we employ the MPS format, which will be introduced in Section~\ref{sec: Matrix product states and tensor trains}. From the corresponding wave functions, we then want to calculate the probability distribution for measuring each basis state as a tensor network. This will be discussed in Section~\ref{sec: Probability distributions and generative sampling}.

\begin{Example}\label{ex: qubit states}
 Every tensor product of $n$ qubits is an $n$-qubit state, e.g.,
 \begin{equation*}
   \begin{split}
    \frac{1}{\sqrt{2}} (\ket{0} - \ket{1}) \otimes \frac{1}{\sqrt{2}} (\ket{0} + \ket{1}) = \frac{1}{2} (\ket{00} + \ket{01} - \ket{10} - \ket{11}) = \frac{1}{2} \begin{bmatrix} 1 \\ -1 \end{bmatrix} \otimes \begin{bmatrix} 1 \\ 1 \end{bmatrix}.
   \end{split}
 \end{equation*}
 The other direction, however, is not true. That is, not every $n$-qubit state can be written as a tensor product of $n$ qubits. Such states are called entangled. Well-known examples of entangled quantum states are the Bell~\cite{Bell1964}, the Greenberger--Horne--Zeilinger~\cite{Greenberger1989}, and the W states~\cite{Duer2000}:
 \begin{align*}
  \ket{\Phi ^{\pm}} &= \frac{1}{\sqrt{2}}(\ket{00} \pm \ket{11}),
  &\ket{\mathrm{GHZ}} &= \frac{1}{\sqrt{2}}(\ket{0 \dots 0} +\ket{1 \dots 1}), \\
  \ket{\Psi ^{\pm}} &= \frac{1}{\sqrt{2}}(\ket{01} \pm \ket{10}), 
  &\ket{\mathrm{W}} &= \frac{1}{\sqrt{n}}(\ket{10 \dots 0} +  \dots + \ket{0 \dots 01}).
 \end{align*}

\end{Example}

\subsection{Quantum gates}
\label{sec: Quantum gates}

\begin{figure}[htbp]
  \centering
  \hfill
  \begin{subfigure}[c]{.125\textwidth}
    \centering
    \begin{quantikz}[row sep={0.3cm,between origins}]
     & & \\
     & & \\
     & & \\
     & \gate{H}&\qw \\
     & & \\
     & & \\
     & &
    \end{quantikz}
    \caption{}
  \end{subfigure}%
  \hfill
  \begin{subfigure}[c]{.125\textwidth}
    \centering
    \begin{quantikz}[row sep={0.3cm,between origins}]
     & & \\
     & & \\
     & & \\
     & \gate{R_\varphi}&\qw \\
     & & \\
     & & \\
     & &
    \end{quantikz}
    \caption{}
  \end{subfigure}%
  \hfill
  \begin{subfigure}{.125\textwidth}
    \centering
    \begin{quantikz}[row sep={0.3cm,between origins}]
     & & \\
     & & \\
     & \ctrl{2}&\qw \\
     & & \\
     & \gate{R_\varphi}&\qw \\
     & & \\
     & & 
    \end{quantikz}
    \caption{}
  \end{subfigure}%
  \hfill
  \begin{subfigure}{.125\textwidth}
    \centering
    \begin{quantikz}[row sep={0.3cm,between origins}]
     & & \\
     & & \\
     & \ctrl{2}&\qw \\
     & & \\
     & \targ{}&\qw \\
     & & \\
     & & 
    \end{quantikz}
    \caption{}
  \end{subfigure}%
  \hfill
  \begin{subfigure}{.125\textwidth}
    \centering
    \begin{quantikz}[row sep={0.3cm,between origins}]
     & & \\
     & \ctrl{4}&\qw \\
     & & \\
     & \control{} & \qw \\
     & & \\
     & \targ{}&\qw \\
     & &
    \end{quantikz}
    \caption{}
  \end{subfigure}%
  \hfill\,
  \caption{Quantum logic gates: (a) Hadamard gate, (b) phase-shift gate, (c) controlled phase-shift gate, (d) CNOT gate, (e) CCNOT gate. Each line depicts a qubit. We follow the usual convention of a state traversing from left to right. Controlled gates act on two or more qubits, where the control qubits are indicated by a dot.}
  \label{fig: gates}
\end{figure}
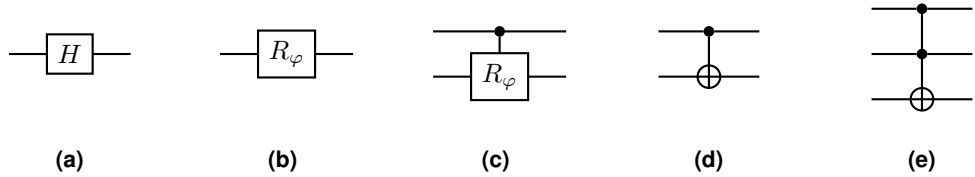

We will now give an overview of essential quantum gates that we will use to construct quantum circuits. 
For a more detailed description and information about other quantum gates, we refer to~\cite{Barenco1995, Nielsen2010}. 
Figure~\ref{fig: gates} shows the considered selection of quantum gates in diagrammatic notation.

Two of the most important single-qubit gates in quantum information processing are the \emph{Hadamard gate} and the \emph{phase-shift gate}, given by the matrix representations
\begin{equation*}
 H = \frac{1}{\sqrt{2}} \begin{bmatrix}1 & 1 \\ 1 &-1\end{bmatrix} \quad \text{and} \quad R_\varphi = \begin{bmatrix} 1 & 0 \\ 0 & e^{i \varphi} \end{bmatrix},
\end{equation*}
respectively. 
See Figure~\ref{fig: gates}~(a) and (b) for the graphical representation of the gates. While the Hadamard gate maps the basis states of a qubit to superpositions of these basis states, a phase-shift gate does not change the probability of measuring $\ket{0}$ or $\ket{1}$ in a fixed basis.
Another essential class of operations is given by (multi-)controlled logic gates, i.e., gates acting on two or more qubits where a specific operation is performed on the target qubits only if certain control qubits are in state $\ket{1}$. 
Examples of controlled gates are the \emph{controlled phase-shift} and the \emph{controlled NOT} gate, see Figure~\ref{fig: gates}~(c) and (d), which can be expressed in canonical format as
\begin{equation}\label{eq: CPHASE gate in CP}
\mathrm{CPHASE}(\varphi)  = \begin{bmatrix} 1 & 0 \\ 0 & 0 \end{bmatrix} \otimes I +  C \otimes R_\varphi = I^{\otimes 2} + C \otimes (R_\varphi-I) \cong \begin{bmatrix} 1 & 0 & 0 & 0 \\ 0 & 1 & 0 & 0 \\ 0 & 0 & 1 & 0 \\ 0 & 0 & 0 & e^{i \varphi} \end{bmatrix}
\end{equation}
and
\begin{equation}\label{eq: CNOT gate in CP}
\mathrm{CNOT}  = \begin{bmatrix} 1 & 0 \\ 0 & 0 \end{bmatrix} \otimes I +  C \otimes \sigma_x = I^{\otimes 2} + C \otimes (\sigma_x-I) \cong \begin{bmatrix} 1 & 0 & 0 & 0 \\ 0 & 1 & 0 & 0 \\ 0 & 0 & 0 & 1 \\ 0 & 0 & 1 & 0 \end{bmatrix},
\end{equation}
respectively. Here, $C$ denotes the control matrix and $\sigma_x$ the Pauli-$X$ gate, given by
\begin{equation*}
 C = \begin{bmatrix} 0 & 0 \\ 0 & 1\end{bmatrix} \quad \text{and} \quad \sigma_x = \begin{bmatrix} 0 & 1 \\ 1 & 0 \end{bmatrix}.
\end{equation*}
Note that for the controlled phase-shift gate the positions of the control and phase-shift matrices can be changed, i.e., $\mathrm{CPHASE} (\varphi) = I^{\otimes 2} + (R_\varphi-I) \otimes C $.

An important three-qubit gate with one target and two control qubits is the \emph{Toffoli} or \emph{CCNOT gate}. Only if both control qubits are in state $\ket{1}$ the NOT operation is carried out. Its matrix and canonical representation are given by
\begin{equation}\label{eq: CCNOT gate in CP}
 \mathrm{CCNOT} = I^{\otimes 3} + C \otimes C \otimes (\sigma_x - I) \cong \begin{bmatrix} 
 1&0&0&0&0&0&0&0\\0&1&0&0&0&0&0&0\\0&0&1&0&0&0&0&0\\0&0&0&1&0&0&0&0\\0&0&0&0&1&0&0&0\\0&0&0&0&0&1&0&0\\0&0&0&0&0&0&0&1\\0&0&0&0&0&0&1&0
 \end{bmatrix}.
\end{equation}
For the sake of completeness, we address another essential type of two-qubit gate, the so-called \emph{SWAP gate} which exchanges the quantum states of two qubits. Although SWAP gates were exploited in previous studies on MPS/MPO-based representations of quantum states~\cite{Banuls2006, Wang2017} and circuits~\cite{Ran2020, Torlai2020}, we will not take SWAP gates explicitly into account. As we will show in Section~\ref{sec: Quantum gates and circuits in TT formalism}, we can construct compact MPO expressions of gates acting on two or more qubits, even if these are not adjacent. 

\subsection{Matrix product states and tensor trains}
\label{sec: Matrix product states and tensor trains}

The storage consumption of a tensor $\T \in \C^D$ with $D=[d_1, \dots, d_n]^\top \in \mathbb{N}^n$ can be estimated as $\mathcal{O}(d^n)$, where $d$ is the maximum of all modes. 
Storing a higher-order tensor is thus in general infeasible for large $n$ since the number of elements of a tensor grows exponentially with the order -- this is also known as the \emph{curse of dimensionality}.
However, it is possible to mitigate this problem by exploiting low-rank tensor approximations.
Our goal is to use the MPS/TT format to efficiently represent quantum states of qubit systems coupled in a one-dimensional chain. 
A tensor $\T \in \C^D$ is said to be in the MPS format if
\begin{equation*}
 \T  = \sum_{k_0=1}^{r_0} \dots \sum_{k_n=1}^{r_n} \bigotimes_{i=1}^n \T^{(i)}_{k_{i-1},:,k_i} = \sum_{k_0=1}^{r_0} \dots \sum_{k_n=1}^{r_n} \T^{(1)}_{k_0,:,k_1} \otimes \dots \otimes \T^{(n)}_{k_{n-1}, :, k_n}.
\end{equation*}
The variables $r_i$ are called \emph{bond dimensions} or \emph{TT ranks} and it holds that $r_0 = r_n = 1$ and $r_i \geq 1$ for $i=1 , \dots, n-1$. 
The tensors $\T^{(i)} \in \C^{r_{i-1} \times d_i \times r_i}$ are called \emph{(TT) cores}. 
Each element of the tensor $\T$ can be written as
\begin{equation}\label{eq: MPS - single entry}
 \T_{x_1, \dots, x_n} = \T^{(1)}_{1,x_1,:} \T^{(2)}_{:,x_2,:} \cdots \T^{(n-1)}_{:,x_{n-1},:} \T^{(n)}_{:,x_n,1},
\end{equation}
which explains the origin of the name matrix product states (MPS). 
Note that here the rank indices start at 1 while the mode indices start at 0. 
In general, every quantum state can be written as an MPS, but the crucial point is that the ranks determine the storage consumption and expressivity.
If the ranks are small enough, we may reduce the storage consumption of an order-$n$ tensor significantly: Instead of an exponential dependence, the storage then depends only linearly on the order and can be estimated as $\mathcal{O}(r^2 d n)$, where $r$ is the maximum over all ranks.
That is, if the underlying correlation structure admits such a low-rank decomposition, an enormous reduction in complexity can be achieved.
As a result, MPSs are well-suited for describing states with weak entanglement.

\begin{figure}[htbp]
\centering
\hfill
\begin{subfigure}[b]{0.45\textwidth}
\centering
\begin{tikzpicture}
\draw[black] (0,0) -- node [label={[shift={(0,-0.15)}]$r_1$}] {} ++ (1,0) ;
\draw[black] (1,0) -- node [label={[shift={(0,-0.15)}]$r_2$}] {} ++ (1,0) ;
\draw[black] (2,0) -- node [label={[shift={(0,-0.15)}]$r_3$}] {} ++ (1,0) ;
\draw[black] (3,0) -- node [label={[shift={(0,-0.15)}]$r_4$}] {} ++ (1,0) ;
\draw[black] (0,0) -- node [label={[shift={(0,-1)}]$d_1$}] {} ++ (0,-0.7) ;
\draw[black] (1,0) -- node [label={[shift={(0,-1)}]$d_2$}] {} ++ (0,-0.7) ;
\draw[black] (2,0) -- node [label={[shift={(0,-1)}]$d_3$}] {} ++ (0,-0.7) ;
\draw[black] (3,0) -- node [label={[shift={(0,-1)}]$d_4$}] {} ++ (0,-0.7) ;
\draw[black] (4,0) -- node [label={[shift={(0,-1)}]$d_5$}] {} ++ (0,-0.7) ;
\node[draw,shape=circle,fill=Blue, scale=0.7] at (0,0){};
\node[draw,shape=circle,fill=Blue, scale=0.7] at (1,0){};
\node[draw,shape=circle,fill=Blue, scale=0.7] at (2,0){};
\node[draw,shape=circle,fill=Blue, scale=0.7] at (3,0){};
\node[draw,shape=circle,fill=Blue, scale=0.7] at (4,0){};
\end{tikzpicture}
\caption{}
\end{subfigure}
\hfill
\begin{subfigure}[b]{0.45\textwidth}
\centering
\begin{tikzpicture}
\draw[black] (0,0) -- node [label={[shift={(0,-0.15)}]$R_1$}] {} ++ (1,0) ;
\draw[black] (1,0) -- node [label={[shift={(0,-0.15)}]$R_2$}] {} ++ (1,0) ;
\draw[black] (2,0) -- node [label={[shift={(0,-0.15)}]$R_3$}] {} ++ (1,0) ;
\draw[black] (3,0) -- node [label={[shift={(0,-0.15)}]$R_4$}] {} ++ (1,0) ;
\draw[black] (0,0) -- node [label={[shift={(0,-1)}]$d_1$}] {} ++ (0,-0.7) ;
\draw[black] (1,0) -- node [label={[shift={(0,-1)}]$d_2$}] {} ++ (0,-0.7) ;
\draw[black] (2,0) -- node [label={[shift={(0,-1)}]$d_3$}] {} ++ (0,-0.7) ;
\draw[black] (3,0) -- node [label={[shift={(0,-1)}]$d_4$}] {} ++ (0,-0.7) ;
\draw[black] (4,0) -- node [label={[shift={(0,-1)}]$d_5$}] {} ++ (0,-0.7) ;
\draw[black] (0,0) -- node [label={[shift={(0,0.2)}]$d_1$}] {} ++ (0,0.7) ;
\draw[black] (1,0) -- node [label={[shift={(0,0.2)}]$d_2$}] {} ++ (0,0.7) ;
\draw[black] (2,0) -- node [label={[shift={(0,0.2)}]$d_3$}] {} ++ (0,0.7) ;
\draw[black] (3,0) -- node [label={[shift={(0,0.2)}]$d_4$}] {} ++ (0,0.7) ;
\draw[black] (4,0) -- node [label={[shift={(0,0.2)}]$d_5$}] {} ++ (0,0.7) ;
\node[draw,shape=circle,fill=Green, scale=0.7] at (0,0){};
\node[draw,shape=circle,fill=Green, scale=0.7] at (1,0){};
\node[draw,shape=circle,fill=Green, scale=0.7] at (2,0){};
\node[draw,shape=circle,fill=Green, scale=0.7] at (3,0){};
\node[draw,shape=circle,fill=Green, scale=0.7] at (4,0){};
\end{tikzpicture}
\caption{}
\end{subfigure}
\hfill
\caption{Graphical representation of the MPS/MPO format:  A core is depicted by a circle with different arms indicating the modes of the tensor and the rank indices. (a) Tensor of order 5 as MPS with ranks $r_1, r_2, r_3, r_4$. The first and the last core are matrices, the other cores are tensors of order 3. (b) An MPO of order 10 with ranks $R_1, R_2, R_3, R_4$. The first and the last core are tensors of order 3, the other cores are tensors of order 4.}
\label{fig: tensor trains}
\end{figure}
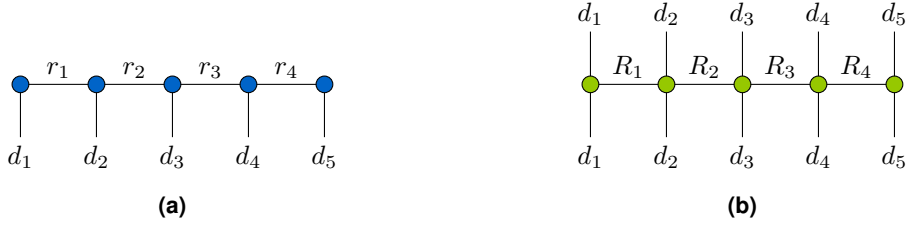

A linear operator $\G \in \C^{D \times D}$ in the MPO/TT format, can be written as
\begin{equation*}
 \G = \sum_{k_0=1}^{R_0} \dots \sum_{k_n=1}^{R_n} \bigotimes_{i=1}^n \G^{(i)}_{k_{i-1},:,:,k_i} = \sum_{k_0=1}^{R_0} \dots \sum_{k_n=1}^{R_n} \G^{(1)}_{k_0,:,:,k_1} \otimes \dots \otimes \G^{(n)}_{k_{n-1}, :, :, k_n}.
\end{equation*}
Here, the cores are tensors of order $4$. 
Figure~\ref{fig: tensor trains}~(a) and (b) show the graphical representation of an MPS $\T \in \C^{D}$ and an MPO $\G \in \C^{D \times D}$ with $D=(d_1, \dots, d_5)$, respectively.

In general, for two tensor operators $\G, \boldH \in \C^{D \times D}$ of order $2 n$ with $D = (2, \dots, 2)$, the product $\G \boldH \in \C^{D \times D}$ is defined as
\begin{equation*}
 (\G \boldH)_{x_1, y_1, \dots, x_n, y_n} = \sum_{z_1 =0}^{1} \dots \sum_{z_n =0}^{1} \G_{x_1, z_1, \dots , x_n , z_n} \boldH_{z_1, y_1, \dots, z_n, y_n},
\end{equation*}
with $x_i, y_i \in \{0,1\}$ for $i=1 , \dots, n$. 
This means that the product of two MPOs $\G$ and $\boldH$ is given by
\begin{equation}\label{eq: gate concatenation}
\begin{split}
 \G \boldH &= \left( \sum_{k_0=1}^{R_0} \dots \sum_{k_n=1}^{R_n} \bigotimes_{i=1}^n \G^{(i)}_{k_{i-1}, :,:, k_i} \right)  \left( \sum_{\ell_0=1}^{R^\prime_0} \dots \sum_{\ell_n=1}^{R^\prime_n} \bigotimes_{i=1}^n \boldH^{(i)}_{\ell_{i-1}, :,:, \ell_i} \right) \\
 &= \sum_{k_0=1}^{R_0} \dots \sum_{k_n=1}^{R_n} \sum_{\ell_0=1}^{R^\prime_0} \dots \sum_{\ell_n=1}^{R^\prime_n} \bigotimes_{i=1}^n \left( \G^{(i)}_{k_{i-1}, :,:, k_i}  \boldH^{(i)}_{\ell_{i-1}, :,:, \ell_i} \right).
\end{split}
\end{equation}
Thus, the ranks of $\G \boldH$ are bounded by the product of the ranks of $\G$ and $\boldH$. 
However, as we will show below, the ranks of resulting circuits can be much smaller in practice. 
The application of an MPO to an MPS, e.g., a quantum circuit and a quantum state, can be seen as a special case of \eqref{eq: gate concatenation}. 
That is, given an MPO $\G \in \C^{D \times D}$ and an MPS $\T \in \C^D$, the product $\G \T$ is given by
\begin{equation*}
\begin{split}
 \G \T &= \left( \sum_{k_0=1}^{R_0} \dots \sum_{k_n=1}^{R_n} \bigotimes_{i=1}^n \G^{(i)}_{k_{i-1}, :,:, k_i} \right)  \left( \sum_{\ell_0=1}^{r_0} \dots \sum_{\ell_n=1}^{r_n} \bigotimes_{i=1}^n \T^{(i)}_{\ell_{i-1}, :, \ell_i} \right) \\
 &= \sum_{k_0=1}^{R_0} \dots \sum_{k_n=1}^{R_n} \sum_{\ell_0=1}^{r_0} \dots \sum_{\ell_n=1}^{r_n} \bigotimes_{i=1}^n \left( \G^{(i)}_{k_{i-1}, :,:, k_i}  \T^{(i)}_{\ell_{i-1}, :, \ell_i} \right).
\end{split}
\end{equation*}

A convenient way of treating MPS tensors is the core notation, cf.~\cite{Kazeev2012, Keller2015, Gelss2017}. We represent the cores as two-dimensional arrays containing either vectors or matrices as elements. For a given MPS $\T \in \C^D$ with cores $\T^{(i)} \in \C^{r_{i-1} \times d_i \times r_i}$, a single core is written as
\begin{equation}
\core{\T^{(i)}} = 
\core{
 \T^{(i)}_{1,:,1} & \cdots & \T^{(i)}_{1,:,r_i} \\
 & & \\
 \vdots & \ddots & \vdots \\
 & & \\
 \T^{(i)}_{r_{i-1},:,1} & \cdots & \T^{(i)}_{r_{i-1},:,r_i}}.
\label{eq: core notation - single core}
\end{equation}
For a given operator $\G \in \C^{D \times D}$ with cores $\G^{(i)} \in \C^{R_{i-1} \times d_i \times d_i \times R_i}$, each core is written as
\begin{equation} \label{eq: core notation - single core - operator}
    \core{\G^{(i)}} =
    \core{
        \G^{(i)}_{1,:,:,1} & \cdots & \G^{(i)}_{1,:,:,R_i} \\
        & & \\
        \vdots & \ddots & \vdots \\
        & & \\
        \G^{(i)}_{R_{i-1},:,:,1} & \cdots & \G^{(i)}_{R_{i-1},:,:,R_i}}.
\end{equation}
We then use the notation
\begin{equation*}
\begin{split}
    \T & = \core{\T^{(1)}} \otimes \dots \otimes \core{\T^{(n)}} \quad\textrm{and}\quad \G  = \core{\G^{(1)}} \otimes \dots \otimes \core{\G^{(n)}}.
    \end{split}
\end{equation*}
The corresponding operations can be regarded as a generalization of the standard matrix multiplication, where the cores contain matrices as elements instead of scalars. Here, we first compute the tensor products of the corresponding elements and then sum over the corresponding columns and rows. The advantage of the core notation is not only that we can derive compact representations of quantum states and logic gates, but that we can also easily manipulate the cores by linear transformations, which is described in Appendix~\ref{app: core manipulation}. 

\begin{Example}\label{ex: W state and CNOT gate as TT}
  Let us consider the W state introduced in Example~\ref{ex: qubit states}. The rank of its canonical representation is $n$ which means that $n$ rank-one tensors (comprising $n$ vectors) are needed to express the amplitude tensor as a sum of tensor products, leading to a storage consumption of $O(2 n^2)$ (without exploiting sparsity). However, the MPS format allows us to represent the wave function of $\ket{\mathrm{W}}$ in a more compact form as
  \begin{equation}\label{eq: W state in TT}
   \PSI =  \frac{1}{\sqrt{n}} \core{\ket{1} & \ket{0}} \otimes \core{\ket{0}&  \\ \ket{1} & \ket{0}} \otimes \dots \otimes \core{\ket{0}&  \\ \ket{1} & \ket{0}}\otimes \core{\ket{0} \\ \ket{1}},
  \end{equation}
  which has a rank of 2, leading to storage consumption of $O(r^2 d n) = O(8n)$.
  For the sake of clarity, we omit core elements that are equal to the vector of all zeros. 
  The decomposition in~\eqref{eq: W state in TT} shows, on the one hand, that MPS representations may require less memory and, on the other hand, that the core notation allows for a more compact representation of qubit states. 
  The form of the decomposition in~\eqref{eq: W state in TT} is a special case of so-called SLIM decompositions used for expressing nearest-neighbor interaction systems in TT format, see~\cite{Gelss2017}. 
  As an example of the core manipulation of MPS/MPO decompositions (see Appendix~\ref{app: core manipulation}), consider the CNOT gate. From~\eqref{eq: CNOT gate in CP}, we deduce its representation in core notation as
  \begin{equation*}
   \mathrm{CNOT} = \core{I & C} \otimes \core{I \\ \sigma_x - I}.
  \end{equation*}
  An alternative MPS representation can be derived as follows:
  \begin{equation*}
  \begin{split}
   \mathrm{CNOT} &= \core{I & C} \otimes \left(\begin{bmatrix} 1 & 0 \\ -1 & 1 \end{bmatrix} \cdot \core{I \\ \sigma_x} \right)\\
   &= \left( \core{I & C} \cdot \begin{bmatrix} 1 & 0 \\ -1 & 1 \end{bmatrix} \right) \otimes  \core{I \\ \sigma_x} \\
   &= \core{I-C & C}  \otimes  \core{I \\ \sigma_x}.
  \end{split}
  \end{equation*}
\end{Example}

\subsection{Probability distributions and generative sampling}
\label{sec: Probability distributions and generative sampling}

Suppose the wave function is given in form of an MPS $\PSI \in \C^{2 ^{\otimes n}}$. 
As explained in~\cite{Ferris2012, Han2018}, it is possible to directly sample from the probability distribution given by the tensor $\mathbf{P}$ with
\begin{equation*}
 \mathbf{P}_{x_1, \dots, x_n} := \mathbb{P}(x) = \left| \PSI_{x_1, \dots, x_n} \right|^2 / Z, 
\end{equation*}
where $Z = \sum_{x_1, \dots, x_n} \left| \PSI_{x_1, \dots, x_n} \right|^2$ is the normalization factor. 
In what follows, we assume that $\PSI$ is already normalized, thus $Z=1$. 
The idea is to exploit the orthonormality of the TT cores to generate samples qubit-wise by constructing conditional probability distributions from segments of the tensor $\mathbf{P}$. 
The tensor $\mathbf{P}$ can be written as
\begin{equation*}
 \PROB_{x_1, \dots, x_n} =  \overline{\PSI_{x_1, \dots, x_n}} \PSI_{x_1, \dots, x_n} \qquad \Longleftrightarrow \qquad \mathbf{P} = \mathrm{diag}(\overline{\PSI}) \PSI ,
\end{equation*}
see Figure~\ref{fig: probability tensors}~(a). Here, $\mathrm{diag}(\,\cdot\,)$ denotes a diagonal MPO, see Appendix~\ref{sec: Diagonal tensors as MPOs}.

\begin{figure}[htbp]
\centering
\begin{subfigure}[b]{0.35\textwidth}
\centering
\begin{tikzpicture}
\draw[black] (0,0) --++ (4,0) ;
\draw[black] (0,1) --++ (4,0) ;
\draw[black] (0,1) --++ (0,-1.75) ;
\draw[black] (1,1) --++ (0,-1.75) ;
\draw[black] (2,1) --++ (0,-1.75) ;
\draw[black] (3,1) --++ (0,-1.75) ;
\draw[black] (4,1) --++ (0,-1.75) ;
\node[draw,shape=circle,fill=Blue, scale=0.7] at (0,1){};
\node[draw,shape=circle,fill=Blue, scale=0.7] at (1,1){};
\node[draw,shape=circle,fill=Blue, scale=0.7] at (2,1){};
\node[draw,shape=circle,fill=Blue, scale=0.7] at (3,1){};
\node[draw,shape=circle,fill=Blue, scale=0.7] at (4,1){};
\node[draw,shape=circle,fill=Green, scale=0.7] at (0,0){};
\node[draw,shape=circle,fill=Green, scale=0.7] at (1,0){};
\node[draw,shape=circle,fill=Green, scale=0.7] at (2,0){};
\node[draw,shape=circle,fill=Green, scale=0.7] at (3,0){};
\node[draw,shape=circle,fill=Green, scale=0.7] at (4,0){};
\end{tikzpicture}
\vspace{0.465cm}
\caption{}%
\end{subfigure}
\hfill
\begin{subfigure}[b]{0.6\textwidth}
\centering
\begin{tikzpicture}
\draw[black] (0,0) --++ (4,0) ;
\draw[black] (0,1) --++ (4,0) ;
\draw[black] (0,1) --++ (0,-1.75) ;
\draw[black] (1,1) --++ (0,-2) ;
\draw[black] (2,1) --++ (0,-1.75) ;
\draw[black] (3,1) --++ (0,-2) ;
\draw[black] (4,1) --++ (0,-1.75) ;
\node[draw,shape=circle,fill=Blue, scale=0.7] at (0,1){};
\node[draw,shape=circle,fill=Blue, scale=0.7] at (1,1){};
\node[draw,shape=circle,fill=Blue, scale=0.7] at (2,1){};
\node[draw,shape=circle,fill=Blue, scale=0.7] at (3,1){};
\node[draw,shape=circle,fill=Blue, scale=0.7] at (4,1){};
\node[draw,shape=circle,fill=Green, scale=0.7] at (0,0){};
\node[draw,shape=circle,fill=Green, scale=0.7] at (1,0){};
\node[draw,shape=circle,fill=Green, scale=0.7] at (2,0){};
\node[draw,shape=circle,fill=Green, scale=0.7] at (3,0){};
\node[draw,shape=circle,fill=Green, scale=0.7] at (4,0){};
\node[draw,shape=circle,fill=white, scale=0.7] at (1,-1){$\mathds{1}$};
\node[draw,shape=circle,fill=white, scale=0.7] at (3,-1){$\mathds{1}$};
\node[] at (5,0){$=$};
\draw[black] (6,0.4) --++ (0,-0.75) ;
\draw[black] (7,0.4) --++ (0,-0.75) ;
\draw[black] (8,0.4) --++ (0,-0.75) ;
\draw[rounded corners=0.12cm,fill=Gray] (5.88, 0.28) rectangle (8.12, 0.52) {};
\end{tikzpicture}
\caption{}
\end{subfigure}\\[0.5cm]
\begin{subfigure}[b]{0.95\textwidth}
\centering
\begin{tikzpicture}
\draw[black] (0,0) --++ (4,0) ;
\draw[black] (0,1) --++ (4,0) ;
\draw[black] (0,1) --++ (0,-2) ;
\draw[black] (1,1) --++ (0,-2) ;
\draw[black] (2,1) --++ (0,-1.75) ;
\draw[black] (3,1) --++ (0,-2) ;
\draw[black] (4,1) --++ (0,-1.75) ;
\node[draw,shape=circle,fill=Blue, scale=0.7] at (0,1){};
\node[draw,shape=semicircle,rotate=225 ,fill=white,inner sep=2pt, anchor=south, outer sep=0pt, scale=0.75] at (1,1){}; 
\node[draw,shape=semicircle,rotate=45,fill=Blue,inner sep=2pt, anchor=south, outer sep=0pt, scale=0.75] at (1,1){};
\node[draw,shape=semicircle,rotate=225 ,fill=white,inner sep=2pt, anchor=south, outer sep=0pt, scale=0.75] at (2,1){}; 
\node[draw,shape=semicircle,rotate=45,fill=Blue,inner sep=2pt, anchor=south, outer sep=0pt, scale=0.75] at (2,1){};
\node[draw,shape=semicircle,rotate=225 ,fill=white,inner sep=2pt, anchor=south, outer sep=0pt, scale=0.75] at (3,1){}; 
\node[draw,shape=semicircle,rotate=45,fill=Blue,inner sep=2pt, anchor=south, outer sep=0pt, scale=0.75] at (3,1){};
\node[draw,shape=semicircle,rotate=225 ,fill=white,inner sep=2pt, anchor=south, outer sep=0pt, scale=0.75] at (4,1){}; 
\node[draw,shape=semicircle,rotate=45,fill=Blue,inner sep=2pt, anchor=south, outer sep=0pt, scale=0.75] at (4,1){};
\node[draw,shape=circle,fill=Green, scale=0.7] at (0,0){};
\node[draw,shape=circle,fill=Green, scale=0.7] at (1,0){};
\node[draw,shape=circle,fill=Green, scale=0.7] at (2,0){};
\node[draw,shape=circle,fill=Green, scale=0.7] at (3,0){};
\node[draw,shape=circle,fill=Green, scale=0.7] at (4,0){};
\node[draw,shape=circle,fill=white, scale=0.7] at (1,-1){$\mathds{1}$};
\node[draw,shape=circle,fill=white, scale=0.7] at (3,-1){$\mathds{1}$};
\node[draw,shape=circle,fill=white, scale=0.7, inner sep=1pt] at (0,-1){\scalebox{0.78}{$\ket{y_1}$}};
\node[draw,shape=circle,fill=white, scale=0.7] at (4,-1){$\mathds{1}$};
\node[] at (5,0){$=$};
\draw[black] (6,0) --++ (4,0) ;
\draw[black] (6,1) --++ (4,0) ;
\draw[black] (6,0) --++ (0,1) ;
\draw[black] (7,0) --++ (0,1) ;
\draw[black] (8,1) --++ (0,-1.75) ;
\draw[black] (9,0) --++ (0,1) ;
\draw[black] (10,0) --++ (0,1) ;
\node[draw,shape=circle,fill=Blue, scale=0.7] at (6,1){};
\node[draw,shape=semicircle,rotate=-45 ,fill=white,inner sep=2pt, anchor=south, outer sep=0pt, scale=0.75] at (7,0){}; 
\node[draw,shape=semicircle,rotate=-225,fill=Blue,inner sep=2pt, anchor=south, outer sep=0pt, scale=0.75] at (7,0){};
\node[draw,shape=semicircle,rotate=225 ,fill=white,inner sep=2pt, anchor=south, outer sep=0pt, scale=0.75] at (8,1){}; 
\node[draw,shape=semicircle,rotate=45,fill=Blue,inner sep=2pt, anchor=south, outer sep=0pt, scale=0.75] at (8,1){};
\node[draw,shape=semicircle,rotate=-45 ,fill=white,inner sep=2pt, anchor=south, outer sep=0pt, scale=0.75] at (9,0){}; 
\node[draw,shape=semicircle,rotate=-225,fill=Blue,inner sep=2pt, anchor=south, outer sep=0pt, scale=0.75] at (9,0){};
\node[draw,shape=semicircle,rotate=-45 ,fill=white,inner sep=2pt, anchor=south, outer sep=0pt, scale=0.75] at (10,0){}; 
\node[draw,shape=semicircle,rotate=-225,fill=Blue,inner sep=2pt, anchor=south, outer sep=0pt, scale=0.75] at (10,0){};
\node[draw,shape=circle,fill=Gray, scale=0.7] at (6,0){};
\node[draw,shape=semicircle,rotate=225 ,fill=white,inner sep=2pt, anchor=south, outer sep=0pt, scale=0.75] at (7,1){}; 
\node[draw,shape=semicircle,rotate=45,fill=Blue,inner sep=2pt, anchor=south, outer sep=0pt, scale=0.75] at (7,1){};
\node[draw,shape=circle,fill=Green, scale=0.7] at (8,0){};
\node[draw,shape=semicircle,rotate=225 ,fill=white,inner sep=2pt, anchor=south, outer sep=0pt, scale=0.75] at (9,1){}; 
\node[draw,shape=semicircle,rotate=45,fill=Blue,inner sep=2pt, anchor=south, outer sep=0pt, scale=0.75] at (9,1){};
\node[draw,shape=semicircle,rotate=225 ,fill=white,inner sep=2pt, anchor=south, outer sep=0pt, scale=0.75] at (10,1){}; 
\node[draw,shape=semicircle,rotate=45,fill=Blue,inner sep=2pt, anchor=south, outer sep=0pt, scale=0.75] at (10,1){};
\node[] at (11,0){$=$};
\draw[black] (13,1) --++ (0,-1.75) ;
\draw[black] (13,0) --++ (-1,0) --++ (0,1) --++ (1,0);
\draw[black] (13,0) --++ (1,0) --++ (0,1) --++ (-1,0);
\node[draw,shape=circle,fill=white, scale=0.7, inner sep = 2.8pt] at (12,0.5){$\Theta$};
\node[draw,shape=circle,fill=white, scale=0.7, inner sep = 3.3pt] at (14,0.5){$I$};
\node[draw,shape=semicircle,rotate=225 ,fill=white,inner sep=2pt, anchor=south, outer sep=0pt, scale=0.75] at (13,1){}; 
\node[draw,shape=semicircle,rotate=45,fill=Blue,inner sep=2pt, anchor=south, outer sep=0pt, scale=0.75] at (13,1){};
\node[draw,shape=circle,fill=Green, scale=0.7] at (13,0){};
\end{tikzpicture}
\caption{}
\end{subfigure}
\caption{Probability distributions and generative sampling: (a) Probability tensor $\mathbf{P}$. Green circles represent the cores of $\mathrm{diag}(\overline{\PSI})$ and blue circles the cores of $\PSI$. (b) Direct computation of the marginal probability tensor $\mathbf{P}^I$ in full format. (c) Construction of conditional probability vector $P^i$ for generative sampling. Here, all cores of $\PSI$ except the first are right-orthonormalized (half-filled circles) before constructing the network. The cores are orthonormal with respect to the modes at the blank halves of the half-filled circles.}
\label{fig: probability tensors}
\end{figure}
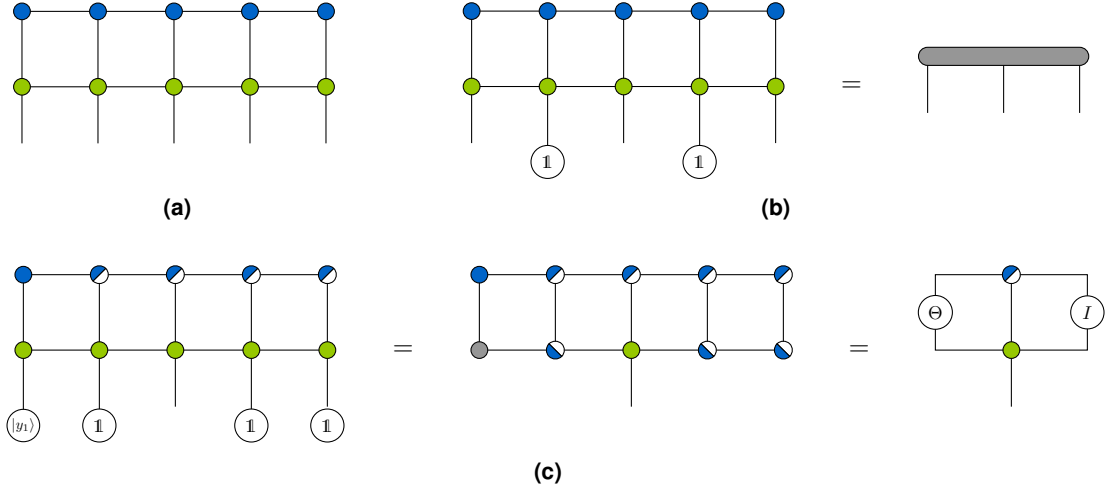

Provided that the number of qubits is small enough so that it can be stored in the full format, one can directly construct the marginal probability tensor corresponding to the measured qubits. That is, by summing the joint probability distribution over all values of the unmeasured qubits, i.e., multiplying the corresponding modes with the vector $\mathds{1} = [1,1]^\top$, and contracting all cores, we obtain a tensor
\begin{equation*}
 \mathbf{P}^I_{x_{i_1}, \dots, x_{i_m}} = \sum_{x_j, j \notin I} \mathbf{P}_{x_{1}, \dots, x_{n}}
\end{equation*}
where $I = \{i_1, \dots, i_m\} \subseteq \{1, \dots, n\}$, with $m \leq n$, denotes the index set of the qubits to be measured. See Figure~\ref{fig: probability tensors}~(b) for a graphical representation of $\mathbf{P}^I$.
If the number of qubits is large, it is impossible to compute the (marginal) probability tensor in the full format. Additionally, we are here rather interested in a generative sampling strategy of the outputs of a given quantum circuit. 
Given the tensor $\PSI$ as MPS, we first right-orthonormalize the last $n-1$ cores, e.g., by applying a sequence of singular value decompositions (SVDs), cf.~Appendix~\ref{app: orthonormalization}. 
The sampling procedure depicted in Figure~\ref{fig: probability tensors}~(c) then works as follows: Suppose we want to measure the $i$th qubit of our register. 
To do so, we sum over all modes with index $j > i$. Modes with index $j < i$ are either contracted with a unit basis vector $\ket{y_j}$ (i.e., $y_j$ was measured before at position $j$) or also multiplied by $\mathds{1}$ (i.e., the $j$th qubit is not measured and, thus, not considered for sampling). 
Note that summation of a mode in this case results in the conjugate transpose of the corresponding right-orthonormal core of $\PSI$. 
Therefore, the segment of the tensor network to the right of the $i$th core of $\mathbf{P}$ is equal to an identity matrix. 
The left part can be contracted iteratively so that we obtain a matrix $\Theta$ at each step of the sampling. 
Given an index set $I$ as above, the resulting vector at position $i_k \in I$, $1 \leq k \leq m$, is then given by
\begin{equation*}
 P^{i_k} = \sum_{x_{i_{k+1}}, \dots, x_{i_m}} \mathbf{P}^I_{y_{i_1}, \dots, y_{i_{k-1}}, :, x_{i_{k+1}}, \dots, x_{i_m}}
\end{equation*}
and contains the conditional probabilities for measuring $x_i$ as the $i$th qubit, given that the states $y_{i_1}, \dots, y_{i_{k-1}}$ are measured on the qubits $i_1, \dots, i_{k-1}$. 
By successively drawing the bit values from the conditional probabilities in $P^{i}$, $i \in I$, we generate a sample according to the probability distribution given by $\mathbf{P}$.

\begin{Remark}
 The process of postselection, i.e., conditioning on the outcome of a measurement on certain qubits, means to condition the probability distribution given by $\mathbf{P}$. That is, after postselecting, e.g., $q_1, \dots, q_p$, we consider the probabilities
 \begin{equation*}
  \mathbb{P} (x_{p+1}= y_{p+1}, \dots, x_n = y_n \mid x_1 = y_1, \dots, x_p = y_p) = \frac{\mathbf{P}_{y_1, \dots, y_n}}{\mathbf{P}^I_{y_1, \dots, y_p}},
 \end{equation*}
 where $I = \{1, \dots, p\}$. The outcomes when measuring $q_{p+1}, \dots, q_n$ then strictly obey the marginal probability distribution given by $\mathbf{P}^J$ with $J = \{p+1, \dots, n\}$ since
 \begin{equation*}
  \sum_{\substack{x_1, \dots , x_p\\ \PROB^I_{x_1, \dots, x_p} > 0}} \mathbf{P}^I_{x_1, \dots, x_p} \frac{\mathbf{P}_{x_1, \dots, x_n}}{\mathbf{P}^I_{x_1, \dots, x_p}} = \sum_{\substack{x_1, \dots , x_p\\ \PROB^I_{x_1, \dots, x_p} > 0}} \mathbf{P}_{x_1, \dots , x_n} = \sum_{x_1, \dots , x_p} \mathbf{P}_{x_1, \dots , x_n} = \mathbf{P}^J_{x_{p+1}, \dots, x_n}.
 \end{equation*}
\end{Remark}

\noindent Given a right-orthonormalized MPS $\PSI$, the computational complexity of the sampling procedure is mainly determined by the contractions of the tensor cores. 
If we store the matrix $\Theta$, see Figure~\ref{fig: probability tensors}~(c), and adapt it successively after each sampling step, the overall cost is $O(s n r^3)$, where $s$ denotes the number of samples and $r$ the maximum rank of $\PSI$.

\section{Representation of quantum gates and circuits as MPOs}
\label{sec: Quantum gates and circuits in TT formalism}

We will now explain how to derive explicit MPO representations of quantum circuits. 
In particular, we will use the core notation of the TT format, see Section~\ref{sec: Matrix product states and tensor trains}, which allows us to derive compact expressions of (interconnected) quantum gates.
Considering a single-qubit gate $A$, e.g., a Hadamard gate, acting on an $n$-qubit system, the tensor operator $\G$ corresponding to the application of $A$ to the qubit at position $p$ is given by
\begin{equation*}
\G = I^{\otimes (p-1)} \otimes A \otimes I^{\otimes (n-p)} = I \otimes \dots \otimes I \otimes \underbrace{A}_{\mathclap{\text{position}~p}} \otimes I \otimes \dots \otimes I.
\end{equation*}
Any controlled gate acting on a quantum register with $n$ qubits has a rank-$2$ representation of the form
\begin{equation*}
\begin{split}
\G &=I^{\otimes n} + I^{\otimes (p-1)} \otimes C \otimes I^{\otimes (q-p-1)} \otimes (A - I) \otimes I^{\otimes (n-q)}\\
&= \core{I} \otimes \dots \otimes \core{I} \otimes \underbrace{\core{ I & C }}_{\mathclap{\text{position}~p}} \otimes \core{I &  \\  & I } \otimes \dots \otimes \core{I &  \\  & I } \otimes \underbrace{\core{I \\ A - I }}_{\mathclap{\text{position}~q}} \otimes \core{I} \otimes \dots \otimes \core{I},
\end{split}
\end{equation*}
where $p$ is the position of the control qubit and $q$ the position of the target qubit.
Again, we here use blanks for core elements equal to the matrix of all zeros.
If $p>q$, the corresponding matrices $C$ and $A-I$ are interchanged. 
If we apply a three-qubit gate, e.g., a \textrm{CCNOT} gate~\eqref{eq: CCNOT gate in CP}, to an $n$-qubit system, we obtain the augmented decomposition
\begin{equation*}
\begin{split}
\G &= \core{I} \otimes \dots \otimes \core{I} \otimes \underbrace{\core{ I & C }}_{\mathclap{\text{position}~p_1}} \otimes \core{I &  \\  & I } \otimes \dots \otimes \core{I &  \\  & I } \otimes \underbrace{\core{I & \\ &  C }}_{\mathclap{\text{position}~p_2}} \otimes \dots\\
& \quad \otimes \core{I &  \\  & I } \otimes \dots \otimes \core{I &  \\  & I } \otimes \underbrace{\core{I \\ A - I }}_{\mathclap{\text{position}~q}} \otimes\core{I} \otimes \dots \otimes \core{I},
\end{split}
\end{equation*}
where $p_1$ and $p_2$ are the positions of the control qubits and $q$ the position of the target qubit.
Note that any (multi-)control quantum gate can be represented by a tensor with canonical and MPS ranks bounded by $2$.

In Sections~\ref{sec: Quantum full adder}--\ref{sec: Quantum Fourier transform}, we will consider specific examples of well-known quantum circuits and express them as compact MPOs resulting from the multiplication/concatenation of several quantum gates. Numerical experiments pertaining to the application of these circuits to quantum states in MPS/TT format can be found in Section~\ref{sec: Numerical experiments}. 

\subsection{Quantum full adder}
\label{sec: Quantum full adder}

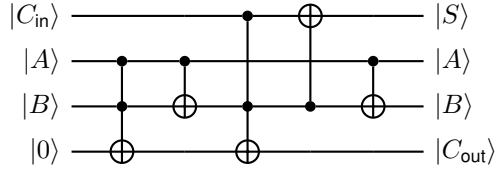
\begin{figure}[htbp]
  \centering
  \begin{quantikz}[row sep={0.6cm,between origins}]
    &\lstick{$\ket{C_\text{in}}$} & \qw        & \qw      & \ctrl{3}   & \targ{}   & \qw      & \rstick{$\ket{S}$}\qw \\
    &\lstick{$\ket{A}$}           & \ctrl{2}   & \ctrl{1} & \qw        & \qw       & \ctrl{1} & \rstick{$\ket{A}$}\qw \\
    &\lstick{$\ket{B}$}           & \control{} & \targ{}  & \control{} & \ctrl{-2} & \targ{}  & \rstick{$\ket{B}$}\qw \\
    &\lstick{$\ket{0}$}           & \targ{}    & \qw      & \targ{}    & \qw       & \qw      & \rstick{$\ket{C_\text{out}}$}\qw
  \end{quantikz}
  \caption{Quantum full adder: Here, the QFA consists of three CNOT and two CCNOT gates.}
  \label{fig: QFA}
\end{figure}

The \emph{quantum full adder} (QFA) is the quantum analogue of a full-adder circuit used in classical computers to add up to three bits. Due to reversibility requirements, the QFA acts on four qubits: The input qubits are $\ket{C_\mathrm{in}}$, $\ket{A}$, $\ket{B}$, and $\ket{0}$ and the output qubits are $\ket{S}$, $\ket{A}$, $\ket{B}$, and $\ket{C_\mathrm{in}}$. Figure~\ref{fig: QFA} shows the QFA implementation proposed in~\cite{Feynman1986}. 

The qubit $\ket{C_\mathrm{in}}$ is carried in from the previous (less-significant) stage of a multi-digit addition. The circuit produces the sum of the input qubits including a carry-out signal for the overflow into the next digit. The corresponding operator is given by the concatenation
\begin{equation*}
 \G = \CNOT(2 \mid 3) \cdot \CNOT(3 \mid 1) \cdot \CCNOT(1,3 \mid 4) \cdot \CNOT(2 \mid 3) \cdot \CCNOT(2,3 \mid 4)  ,
\end{equation*}
where $\CNOT(p \mid q)$ and $\CCNOT(p_1, p_2 \mid q)$ denote the (controlled-)controlled NOT gates with control qubits $p, p_1, p_2$ and target qubit $q$. That is, using the MPO representation of the involved gates, we consider the product
\begin{equation}\label{eq: QFA product}
\begin{array}{ c @{} c @{} c @{} c @{} c @{} c @{} c @{} c @{} c}
\G \, = & ~& I & ~\otimes~ & \core{ I & C } & ~\otimes~ & \core{ I \\ \sigma_x - I  } & ~ \otimes ~ & I\\[0.3cm]
& \hspace*{0.25cm}\cdot\hspace*{0.25cm} & \core{ I & \sigma_x - I } & \otimes & \core{ I &  \\  & I } & \otimes & \core{ I \\ C  } & \otimes & I\\[0.3cm]
& \cdot & \core{ I & C } & \otimes & \core{ I &  \\  & I } & \otimes & \core{ I & \\  & C  } & \otimes & \core{ I \\ \sigma_x - I }\\[0.3cm]
& \cdot & I & \otimes & \core{ I & C } & \otimes & \core{ I \\ \sigma_x - I  } & \otimes & I\\[0.3cm]
& \cdot & I & \otimes & \core{ I & C } & \otimes & \core{ I &  \\  & C  } & \otimes & \core{ I \\ \sigma_x -I }\mathclap{.}
\end{array}
\end{equation}
Although the theoretical rank bounds of an MPO decomposition of an operator in $\C^{(2 \times 2)^{\times 4}}$ are given by $(1,4,16,4,1)$, we are able to provide an exact representation with much smaller ranks. In fact, the operator $\G$ can be written as
\begin{equation}\label{eq: QFA MPO}
 \G = \core{\sigma_x C_0 & I & \sigma_x C_1 } \otimes \core{C_0 & C_1 &  &  \\  & C_0 & C_1 &  \\ &  & C_0 & C_1} \otimes \core{C_1 &  \\ C_0 &  \\  & C_1 \\  & C_0} \otimes \core{I \\ \sigma_x},
\end{equation}
with control matrices $C_0 = I - C$ and $C_1 = C$.
A proof of correctness of \eqref{eq: QFA MPO} can be found in Appendix~\ref{app: QFA}. 
The coupling between the different dimensions directly reflects the action of the QFA on different basis states.
For instance, if both qubits $\ket{A}$ and $\ket{B}$ are in state $\ket{0}$, we have
\begin{equation*}
\begin{split}
 \ket{S, A, B, C_\mathrm{out}} &=\G \ket{C_\mathrm{in}, A, B, 0}\\
 &= \core{\sigma_x C_0 \ket{C_\mathrm{in}} & \ket{C_\mathrm{in}} & \sigma_x C_1 \ket{C_\mathrm{in}} } \otimes \core{\ket{0} & 0 & 0 & 0 \\ 0 & \ket{0} & 0 & 0 \\ 0& 0 & \ket{0} & 0} \otimes \core{0 & 0 \\ \ket{0} & 0 \\ 0 & 0 \\ 0 & \ket{0}} \otimes \core{\ket{0} \\ \ket{1}}\\
 &= \core{\sigma_x C_0 \ket{C_\mathrm{in}} & \ket{C_\mathrm{in}} & \sigma_x C_1 \ket{C_\mathrm{in}} } \otimes \core{0 & 0\\ \ket{0,0} &0\\0 & 0 } \otimes \core{\ket{0} \\ \ket{1}}\\
 &= \ket{C_\mathrm{in},0,0,0}.
\end{split}
\end{equation*}
Analogously, we get $\ket{S, A, B, C_\mathrm{out}} = \ket{C_\mathrm{in},1,1,1}$ when $\ket{A}=\ket{B}=\ket{1}$.
In Section Section~\ref{sec: Quantum full adder network}, we will use the carry-in and carry-out qubits to couple several QFAs.

\subsection{Simon's algorithm}
\label{sec: Simons algorithm}

\begin{figure}[htbp]
  \centering
  \begin{subfigure}[t]{0.47\textwidth}
   \centering
     \begin{quantikz}[row sep={0.6cm,between origins}]
        &\lstick{$\ket{0}_1$} & \gate{H} & \gate[8, nwires={3,7}]{U_f} & \qw & \gate{H}  &  \meter{} \\
        &\lstick{$\ket{0}_1$} & \gate{H} &        &  \qw     & \gate{H}  &  \meter{} \\[-0.1cm]
        &                   & \vdots   &         &      & \vdots    & \\[0.1cm]
        &\lstick{$\ket{0}_1$} & \gate{H} &       &   \qw     & \gate{H}  &  \meter{} \\
        &\lstick{$\ket{0}_2$} & \qw & & \meter{} &  \qw & \qw\\
        &\lstick{$\ket{0}_2$} & \qw & & \meter{} & \qw & \qw \\[-0.1cm]
        &                   & \vdots   &               & \vdots    & \\[0.1cm]
        &\lstick{$\ket{0}_2$} &  \qw & & \meter{} & \qw & \qw \\
    \end{quantikz}%
    \caption{}
  \end{subfigure}%
  \hfill
  \begin{subfigure}[t]{0.5\textwidth}
   \centering
     \begin{quantikz}[row sep={0.6cm,between origins}]
        &\lstick{$\ket{0}_1$} & \gate{H} & \gate[7, nwires={5}]{U^\prime_f} & \qw & \gate{H}  &  \meter{} \\
        &\lstick{$\ket{0}_2$} & \qw      &        &  \meter{}     & \qw  &  \qw \\
        &\lstick{$\ket{0}_1$} & \gate{H} &        & \qw & \gate{H}  &  \meter{} \\
        &\lstick{$\ket{0}_2$} & \qw      &        &  \meter{}     & \qw       &  \qw \\[0.08cm]
        &                     & \vdots   &        &               & \vdots    & \\[0.45cm]
        &\lstick{$\ket{0}_1$} & \gate{H} &        & \qw           & \gate{H}  &  \meter{} \\
        &\lstick{$\ket{0}_2$} & \qw      &        &  \meter{}     & \qw       &  \qw \\
    \end{quantikz}%
    \caption{}
  \end{subfigure}\\[0.5cm]
  \begin{subfigure}[t]{\textwidth}
   \centering
     \begin{quantikz}[row sep={0.6cm,between origins}]
        \lstick{$\ket{0}_1$} & \gate{H}\gategroup[8,steps=1,style={draw=none}, background, label style={label position=below,anchor=north,yshift=-0.2cm}]{$\G_{1}$}\slice[style=black]{} & [1cm]\ctrl{1}\gategroup[8,steps=4,style={draw=none}, background, label style={label position=below,anchor=north,yshift=-0.2cm}]{$\G_{2}$} & \qw      & \qw      & \qw\slice[style=black]{}      & [1cm]\ctrl{1}\gategroup[8,steps=2,style={draw=none}, background, label style={label position=below,anchor=north,yshift=-0.2cm}]{$\G_{3}$} & \ctrl{5}\slice[style=black]{} & [1cm]\gate{H}\gategroup[8,steps=1,style={draw=none}, background, label style={label position=below,anchor=north,yshift=-0.2cm}]{$\G_{4}$}\slice[style=black]{}  & [1cm]\qw &  \meter{} \\
        \lstick{$\ket{0}_2$} & \qw      & \targ{}  & \qw      & \qw      & \qw      & \targ{}  & \qw      & \qw  &  \meter{} & \qw \\
        \lstick{$\ket{0}_1$} & \gate{H} & \qw      & \ctrl{1} & \qw      & \qw      & \qw      & \qw      & \gate{H}  & \qw &  \meter{} \\
        \lstick{$\ket{0}_2$} & \qw      & \qw      & \targ{}  & \qw      & \qw      & \qw      & \qw      & \qw       &  \meter{} & \qw \\
        \lstick{$\ket{0}_1$} & \gate{H} & \qw      & \qw      & \ctrl{1} & \qw      & \qw      & \qw      & \gate{H}  & \qw &  \meter{} \\
        \lstick{$\ket{0}_2$} & \qw      & \qw      & \qw      & \targ{}  & \qw      & \qw      & \targ{}  & \qw       &  \meter{} & \qw \\
        \lstick{$\ket{0}_1$} & \gate{H} & \qw      & \qw      & \qw      & \ctrl{1} & \qw      & \qw      & \gate{H}  & \qw &  \meter{} \\
        \lstick{$\ket{0}_2$} & \qw      & \qw      & \qw      & \qw      & \targ{}  & \qw      & \qw      & \qw       &  \meter{} & \qw 
    \end{quantikz}%
    \caption{}
  \end{subfigure}%
  \caption{Simon's circuit: (a) Diagrammatic notation of Simon's algorithm. The qubits $\ket{0}_1$ and $\ket{0}_2$ belong to the first and second register, respectively. (b) Reordering of the quantum registers so that the qubits alternate between the first and the second register. (c) Example of Simon's algorithm involving an oracle based on CNOT gates. Since the Hadamard gates on the first register and the measurements on the last register act on disjoint sets of qubits and thus commute, we can change the order of these operations.}
  \label{fig: Simon}
\end{figure}
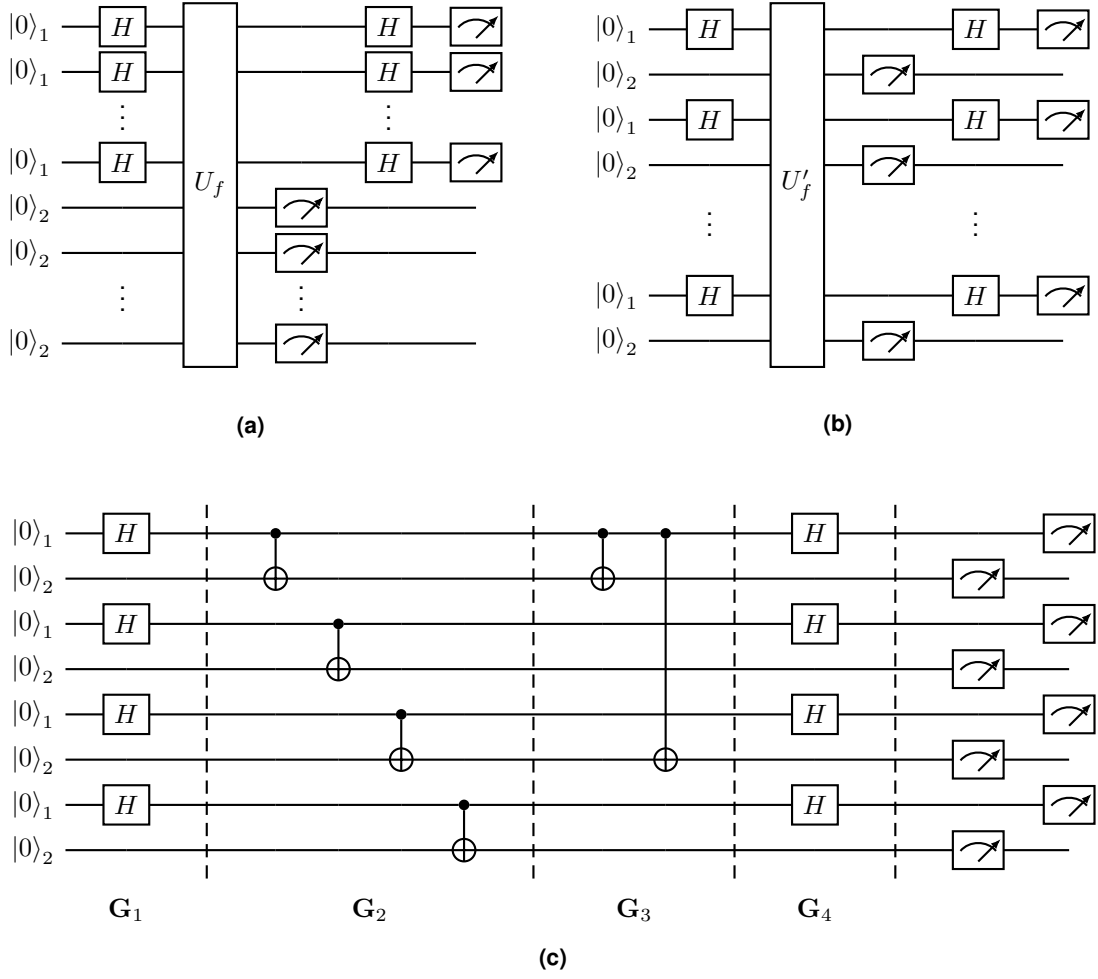

We begin our considerations with the first quantum algorithm that demonstrated an exponential speed-up compared to the best classical algorithm, namely \emph{Simon's algorithm}~\cite{Simon1997, Johansson2017}. Let $f \colon \{0,1\}^n \to \{0,1\}^n$ be a function that is either \emph{one-to-one} (maps exactly one input to every unique output) or \emph{two-to-one} (maps exactly two inputs to every unique output) such that 
\begin{equation*}
 f(x) = f(y) \quad \Longleftrightarrow \quad x = y \oplus b
\end{equation*}
for all $x,y \in \{0,1\}^n$, where $\oplus$ denotes bitwise XOR. The vector $b \in \{0,1\}^n$ is called a \emph{hidden bitstring} and the aim is to determine $b$ (also called \emph{Simon's problem}) minimizing the number of evaluations of $f$. Even though Simon's algorithm assumes the existence of a potentially highly complex black-box oracle $U_f$ for the function evaluation of $f$, see below, it was an important step in quantum computing as it solves Simon's problem exponentially faster and with exponentially fewer queries than the best classical algorithm and served as an inspiration for Shor's algorithm, see Section~\ref{sec: Shor's algorithm}. It was shown recently that Simon's algorithm can also be used to attack certain types of cryptosystems~\cite{Santoli2017, Shinagawa2022}. The method involves the following steps:
\begin{enumerate}
 \item[1)] Initialize two $n$-qubit registers with the zero state, the first one as input and the second one as output of the oracle.
 \item[2)] Apply Hadamard gates to the first register to create a superposition of all possible inputs.
 \item[3)] Apply the oracle function so that $\ket{x,0^n} \mapsto \ket{x, f(x)}$. 
 \item[4)] Measure the second register, $f(x)$, thus collapse a superposition over the first register, i.e., $\frac{1}{\sqrt{2}} (\ket{x} + \ket{x \oplus b})$.
 \item[5)] Apply Hadamard gates to the first register.
 \item[6)] Measure the first register and obtain a bitstring $z$ with bitwise inner product $z \cdot b = 0~(\bmod~{2})$.
\end{enumerate}
The corresponding quantum circuit is shown in Figure~\ref{fig: Simon}~(a). The postprocessing step, i.e., determining the hidden bitstring $b$ from measurements $z_1, \dots, z_m$, can then be realized on a classical computer by solving a system of linear equations with mod-$2$ arithmetic, where a solution $b=00 \dots 0$ represents the case of a one-to-one function $f$.

First, we reorder the qubits in both registers, see Figure~\ref{fig: Simon}~(b), so that a low-rank decomposition of the circuit can be constructed. Let us consider the oracle $U^\prime_f$ shown in Figure~\ref{fig: Simon}~(c) acting on $2\cdot4 = 8$ qubits. By construction, the associated gate groups $\G_2$ and $\G_3$ encode the hidden bitstring $b=1010$ where $\G_2$ copies the content of the first register to the second register and $\G_3$ flips the first and third qubit of the second register if the leading qubit of the first register is in state $1$. This becomes clear by considering the following mappings which show that the oracle indeed satisfies $f(x) = f(x \oplus b)$:
\begin{equation}\label{eq: Simon - mappings}
 0 i_2 0 i_4 \mapsto 0 i_2 0 i_4, \quad 0 i_2 1 i_4 \mapsto 0 i_2 1 i_4, \quad 1 i_2 0 i_4 \mapsto 0 i_2 1 i_4, \quad 1 i_2 1 i_4 \mapsto 0 i_2 0 i_4.
\end{equation}
Defining $C_0 = I - C$ and $C_1 =C$ again, it follows that $\G_2$ can be represented as
\begin{equation*}
\begin{split}
 \G_2 &= (C_0 \otimes I + C_1 \otimes \sigma_x) \otimes (C_0 \otimes I + C_1 \otimes \sigma_x) \otimes (C_0 \otimes I + C_1 \otimes \sigma_x) \otimes (C_0 \otimes I + C_1 \otimes \sigma_x)\\
 &= \core{C_0 & C_1} \otimes \core{I \\ \sigma_x} \otimes \core{C_0 & C_1} \otimes \core{I \\ \sigma_x} \otimes \core{C_0 & C_1} \otimes \core{I \\ \sigma_x} \otimes \core{C_0 & C_1} \otimes \core{I \\ \sigma_x},
\end{split}
\end{equation*}
and the whole circuit $\G= \G_4 \cdot \G_3 \cdot \G_2 \cdot \G_1$ has an MPO representation in the form of
\begin{equation}\label{eq: Simon - gates}
 \begin{split}
 \G&= \core{ A & B } \otimes \core{ I & \\  & I } \otimes \core { A & B &  & \\  &  & A & B } \otimes \core{ I &  \\ \sigma_x &  \\  & I \\  & \sigma_x } \\
 & \quad \otimes \core{ A & B \\ B & A } \otimes \core{ I \\ \sigma_x } \otimes \core{ A & B  } \otimes \core{ I \\ \sigma_x },
 \end{split}
\end{equation}
where $A = H C_0 H = \frac{1}{2} \begin{bmatrix} 1 & 1 \\ 1 & 1\end{bmatrix}$ and $B = H C_1 H = \frac{1}{2} \begin{bmatrix} 1 & -1 \\ -1 & 1\end{bmatrix}$. See Appendix~\ref{app: Simon} for a detailed derivation of the above expressions.
Note that due to the reordering of the qubits we can write $\G$ as an MPO with ranks bounded by 4.
Furthermore, the NOT operation on the second qubit vanishes since the CNOT gates in $\G_2$ and $\G_3$ acting on the first two qubits cancel each other out.
The application of the operator given in~\eqref{eq: Simon - gates} to the zero state yields the quantum state
\begin{equation*}
 \begin{split}
  \G \ket{0^{2n}} &= \G \cdot \left( \begin{bmatrix} 1 \\ 0 \end{bmatrix} \otimes \begin{bmatrix} 1 \\ 0 \end{bmatrix} \otimes \dots \otimes \begin{bmatrix} 1 \\ 0 \end{bmatrix} \right) \\
  &= \frac{1}{4} \cdot  \core{  \ket{+} & \ket{-} } \otimes \core{\ket{0} &  \\[0.5em]  & \ket{0} } \otimes \core{\ket{+} & \ket{-} &  & \\[0.5em]  &  & \ket{+} & \ket{-} } \otimes \core{  \ket{0} &  \\[0.5em] \ket{1} &  \\[0.5em]  & \ket{0} \\[0.5em]  & \ket{1}  } \\
 & \quad \otimes \core{ \ket{+} & \ket{-} \\[0.5em] \ket{-} & \ket{+} } \otimes \core{  \ket{0} \\[0.5em] \ket{1} } \otimes \core{ \ket{+} & \ket{-} } \otimes \core{  \ket{0} \\[0.5em] \ket{1}  }.
 \end{split}
\end{equation*}
where $\ket{+}$ and $\ket{-}$ denote the X-basis states $\frac{1}{\sqrt{2}} [1,1]^\top$ and $\frac{1}{\sqrt{2}} [1,-1]^\top$, respectively. 
As we can directly see, the wave function is always zero if we consider the first bit of the second register (corresponds to the second TT core) to be in state $\ket{1}$. 
This coincides with the possible mappings given in~\eqref{eq: Simon - mappings}. 
We construct the probability tensor $\PROB \in \R^{2^{\times 8}}$ as described in Section~\ref{sec: Probability distributions and generative sampling}. 
As shown above, the postselection on a measurement outcome in the second register, say $j_1 j_2 j_3 j_4$, corresponds to computing the conditional probability tensor $\PROB^C = \PROB_{:,j_1,:,j_2,:,j_3,:,j_4} \in \mathbb {R}^{2^{\times 4}}$. 
That is, the probability of obtaining the bitstring $i_1 i_2 i_3 i_4$ when we measure the first register (after obtaining $j_1 j_2 j_3 j_4$ on the second register) is given by $\PROB^C_{i_1, i_2, i_3, i_4} = \PROB_{i_1, j_1, i_2, j_2, i_3, j_3, i_4, j_4}$. 
Finally, the probability distribution of all possible outcomes of the first register is given by the marginal probability tensor 
\begin{equation*}
\begin{split}
 \PROB^M = \sum_{j_1, j_2 , j_3, j_4 = 1} ^2 \PROB_{:,j_1, :, j_2, :, j_3, : , j_4} = \frac{1}{8} \cdot \core{\ket{0} & \ket{1}} \otimes \core{\ket{+} & \\ & \ket{+}} \otimes \core{\ket{0} \\ \ket{1}} \otimes \core{\ket{+}}.
\end{split}
\end{equation*}
The set of bitstrings with non-zero probability is given by
\begin{equation*}
 L=\{0000, 0001, 0100, 0101, 1010, 1011, 1110, 1111\},
\end{equation*}
each bitstring having a probability of $12.5 \%$, and it turns out that $1010$ is the only bitstring $b \neq 0$ satisfying $z \cdot b = 0~(\bmod~2)$ for each $z \in L$.

\subsection{(Inverse) Quantum Fourier transform}
\label{sec: Quantum Fourier transform}

\begin{figure}[htbp]
  \centering
  \begin{subfigure}[t]{\textwidth}
   \centering
  \begin{quantikz}[column sep=0.3cm, row sep={0.6cm,between origins}]
    \lstick{$\ket{x_1}$}     & \gate{H}\gategroup[6,steps=4,style={draw=none}, background, label style={label position=below,anchor=north,yshift=-0.2cm}]{$\G_1$} & \gate{R_{2}} & \,\ldots\,\qw & \gate{R_n}\slice[style=black]{} & \qw      & \qw        & \qw         & \qw            & \qw         & \qw      & \qw        & \qw      & \rstick{$\ket{\psi_n}$}\qw \\
    \lstick{$\ket{x_2}$}     & \qw      & \ctrl{-1}      & \qw            & \qw        & \gate{H}\gategroup[5,steps=4,style={draw=none}, background, label style={label position=below,anchor=north,yshift=-0.2cm}]{$\G_2$} & \gate{R_2} & \,\ldots\,\qw & \gate{R_{n-1}}\slice[style=black]{} & \qw         & \qw      & \qw        & \qw      & \rstick{$\ket{\psi_{n-1}}$}\qw \\
    \lstick{$\ket{x_3}$}     & \qw      & \qw              & \qw        & \qw   & \qw   & \ctrl{-1}  & \qw         & \qw            & \,\ldots\,\qw & \qw      & \qw        & \qw      & \rstick{$\ket{\psi_{n-2}}$}\qw \\[-0.2cm]
    \lstick{$\vdots$~~~}     &              &                &            &          &            &             &                &             &          &            &     &     & \rstick{~~$\vdots$}\\
    \lstick{$\ket{x_{n-1}}$} & \qw              & \qw         & \qw      & \qw        & \qw      & \qw        & \qw         & \qw            & \qw & \gate{H}\gategroup[2,steps=2,style={draw=none}, background, label style={label position=below,anchor=
north,yshift=-0.2cm}]{$\G_{n-1}$} & \gate{R_2}\slice[style=black]{} & \qw      & \rstick{$\ket{\psi_{2}}$}\qw\\
    \lstick{$\ket{x_n}$}     & \qw            & \qw         & \qw            & \ctrl{-5}  & \qw      & \qw        & \qw         & \ctrl{-4}      & \qw\slice[style=black]{} & \qw      & \ctrl{-1}  & \gate{H}\gategroup[1,steps=1,style={draw=none}, background, label style={label position=below,anchor=
north,yshift=-0.2cm}]{$\G_{n}$} & \rstick{$\ket{\psi_1}$}\qw
  \end{quantikz}
  \caption{}
  \end{subfigure}
  \begin{subfigure}[t]{\textwidth}
   \centering
  \begin{quantikz}[column sep=0.3cm, row sep={0.6cm,between origins}]
    \lstick{$\ket{\psi_n}$}     & \qw\gategroup[6,steps=1,style={draw=none}, background, label style={label position=below,anchor=north,yshift=-0.2cm}]{$\G_n^{-1}$}\slice[style=black]{} & \qw\gategroup[6,steps=2,style={draw=none}, background, label style={label position=below,anchor=north,yshift=-0.2cm}]{$\G_{n-1}^{-1}$} & \qw\slice[style=black]{} & \qw\slice[style=black]{} & \qw\gategroup[6,steps=4,style={draw=none}, background, label style={label position=below,anchor=north,yshift=-0.2cm}]{$\G_{2}^{-1}$} & \qw & \qw & \qw\slice[style=black]{} & \gate{R_{n}^*}\gategroup[6,steps=4,style={draw=none}, background, label style={label position=below,anchor=north,yshift=-0.2cm}]{$\G_{1}^{-1}$} & \,\ldots\,\qw & \gate{R_2^*} & \gate{H} & \rstick{$\ket{\phi_1}$}\qw \\
    \lstick{$\ket{\psi_{n-1}}$} & \qw & \qw & \qw & \qw           & \gate{R_{n-1}^*} & \,\ldots\,\qw & \gate{R_2^*} & \gate{H} & \qw & \qw & \ctrl{-1} & \qw & \rstick{$\ket{\phi_2}$}\qw \\
    \lstick{$\ket{\psi_{n-2}}$} & \qw & \qw & \qw & \,\ldots\,\qw & \qw & \qw & \ctrl{-1} & \qw & \qw & \qw & \qw & \qw & \rstick{$\ket{\phi_3}$}\qw \\[-0.2cm]
    \lstick{$\vdots$~~~} & & & & & & & & & & & & & \rstick{~~$\vdots$}\\
    \lstick{$\ket{\psi_2}$}     & \qw & \gate{R_2^*} & \gate{H} & \qw & \qw & \qw & \qw & \qw & \qw & \qw & \qw & \qw & \rstick{$\ket{\phi_{n-1}}$}\qw\\
    \lstick{$\ket{\psi_1}$}     & \gate{H} & \ctrl{-1} & \qw & \qw & \ctrl{-4} & \qw & \qw & \qw & \ctrl{-5} & \qw & \qw & \qw & \rstick{$\ket{\phi_n}$}\qw
  \end{quantikz}
  \caption{}
  \end{subfigure}
  \caption{(Inverse) quantum Fourier transform: (a) QFT applied to a basis state $\ket{x}$ results in a quantum state $\psi$. To reverse the ordering of the qubits at most $n/2$ would have to be included at the end of the circuit. (b) The application of the inverse QFT to a (reordered) Fourier state yields a quantum state $\ket{\phi}$ in computational basis.}
  \label{fig: QFT}
\end{figure}
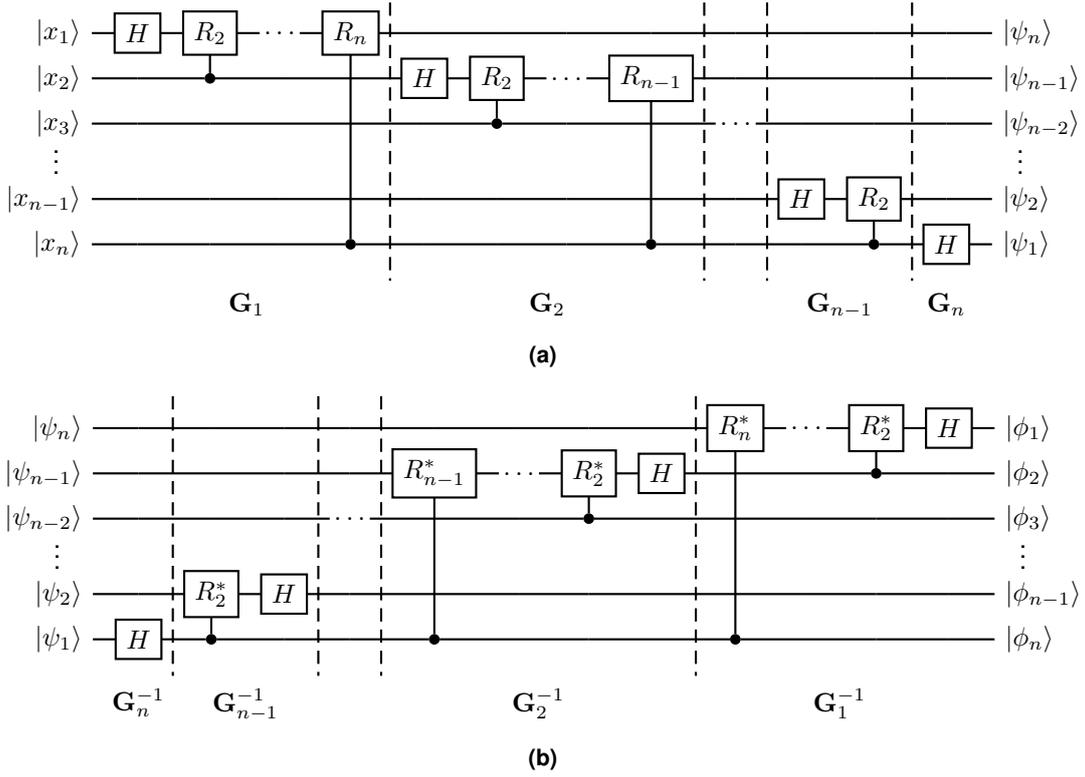

The quantum counterpart of the classical discrete Fourier transform is the \emph{quantum Fourier transform} (QFT)~\cite{Coppersmith1994}, which performs the transformation on the amplitude tensor of a quantum state.
Even compared to the most efficient implementations on classical computers like fast Fourier transform, QFT provides an exponential speed-up on quantum computers.
Due to the fact that the result of the QFT is not directly accessible as a whole, the application areas of the classical and the quantum versions of the Fourier transform differ.
However, QFT is an essential part of various quantum algorithms, in particular Shor's algorithm for finding the prime factors of a given integer~\cite{Shor1994}.
It maps a state in the computational basis to a state in the Fourier basis.
That is, given a basis state $\ket{x} = \ket{x_1 \dots x_n} = \ket{x_1} \otimes \dots \otimes \ket{x_n}$ with $x_j \in \{0,1\}$ for $j=1, \dots,n$, the result after applying the QFT is
\begin{equation}\label{eq: QFT basis state}
 \mathrm{QFT} \ket{x} = \frac{1}{\sqrt{N}} \sum_{y=0}^{N-1} e^{\frac{2 \pi \mathrm{i}}{N} x y}\ket{y} = \frac{1}{\sqrt{N}} \left(\ket{0} + e^{2 \pi \mathrm{i} 0.x_n } \ket{1} \right) \otimes \dots \otimes \left(\ket{0} + e^{2 \pi \mathrm{i} 0.x_1 \dots x_n } \ket{1} \right),
\end{equation}
where $N = 2^n$ and $0.x_n, \dots, 0.x_1 \dots x_n$ denote fractional binary numbers, i.e., $0.y_1 \dots y_m = \sum_{j=1}^m y_j 2^{-j}$.
Figure~\ref{fig: QFT}~(a) shows the corresponding circuit diagram. The quantum circuit can be divided into $n$ gate groups $\G_i$, each composed of Hadamard and controlled phase gates acting on qubit $i$.
The single-qubit gates $R_k$ are defined as the diagonal matrices $ \diag \left(1, e^{\frac{2 \pi \mathrm{i}}{2^k}} \right)$.

\noindent Note that the outputs $\psi_i = \frac{1}{\sqrt{2}} (\ket{0} + e^{2 \pi \mathrm{i} 0.x_i \dots x_n} \ket{1}) $ are in reverse order. However, we omit a reordering (by using SWAP gates) because we can simply read the resulting MPS backwards.
While $\G_n$ is given by $\G_n = I^{\otimes (n-1)} \otimes H$, each gate group $\G_i$ of the QFT with $1 \leq i \leq n-1$ can be represented as
\begin{equation}\label{eq: QFT gate group}
  \G_i = \frac{1}{\sqrt{2}} I^{\otimes(i-1)}  \otimes \core{ \begin{bmatrix}1 & 1 \\ 0 & 0 \end{bmatrix} & \begin{bmatrix}0 & 0 \\ 1 & -1 \end{bmatrix}
 } \otimes \core{I &  \\  & R_2 } \otimes \dots \otimes \core{I &  \\  & R_{n-i}} \otimes \core{I \\ R_{n-i+1}}.
\end{equation}
The derivation of the above decomposition is shown in Appendix~\ref{app: QFT}.
Analog expression in terms of Kronecker products and quantized tensor trains can be found in~\cite{Camps2021} and~\cite{Dolgov2012}, respectively.
From \eqref{eq: QFT basis state} we know that $\mathrm{QFT} \ket{x}$ can be written as a rank-one tensor for any basis state $\ket{x}$. 
Even though the operator in~\eqref{eq: QFT gate group} has rank 2, we can easily show a form of rank preservation which induces the rank-one decomposition of $\mathrm{QFT} \ket{x}$.
Given a quantum state 
\begin{equation*}
\ket{\psi} = \ket{\psi_1} \otimes \dots \otimes \ket{\psi_{i-1}} \otimes \ket{x_i} \otimes \dots \otimes \ket{x_n},
\end{equation*}
where $\psi_1, \dots, \psi_{i-1}$ are arbitrary superpositions and $x_i , \dots, x_n$ are basis states, we have
\begin{equation*}
 \begin{split}
  \G_i \ket{\psi} &= \frac{1}{\sqrt{2}} \ket{\psi_1 \dots \psi_{i-1}} \otimes \core{\ket{0} &  e^{2 \pi \mathrm{i} 0.x_i 00 \dots 000}\ket{1}} \otimes \core{\ket{x_{i+1}} &  \\ & e^{2 \pi \mathrm{i} 0.0 x_{i+1}0 \dots 000} \ket{x_{i+1}}} \otimes \dots\\
  & \qquad \dots \otimes \core{\ket{x_{n-1}} &  \\ & e^{2 \pi \mathrm{i} 0.000 \dots 0 x_{n-1} 0} \ket{x_{n-1}}} \otimes \core{\ket{x_{n}} \\ e^{2 \pi \mathrm{i} 0.000 \dots 0 0 x_{n}} \ket{x_{n}}}\\
  &= \frac{1}{\sqrt{2}} \ket{\psi_1 \dots \psi_{i-1}} \otimes \core{\ket{0} &  e^{2 \pi \mathrm{i} 0.x_i 00 \dots 000}\ket{1}} \otimes \core{\ket{x_{i+1} \dots x_n} \\ e^{2 \pi \mathrm{i} 0.0 x_{i+1} \dots x_n} \ket{x_{i+1} \dots x_n}} \\
  &=  \frac{1}{\sqrt{2}} \ket{\psi_1 \dots \psi_{i-1}} \otimes \left( \ket{0} +  e^{2 \pi \mathrm{i} 0.x_i  \dots x_n}\ket{1} \right) \otimes \ket{x_{i+1} \dots x_n}.
 \end{split}
\end{equation*}
Thus, it holds that
\begin{equation*}
 \G_{i} \cdot \ldots \cdot \G_1 \ket{x} = \frac{1}{\sqrt{2^i}} \left( \ket{0} +  e^{2 \pi \mathrm{i} 0.x_1  \dots x_n}\ket{1} \right) \otimes  \dots \otimes \left( \ket{0} +  e^{2 \pi \mathrm{i} 0.x_i  \dots x_n}\ket{1} \right) \otimes \ket{x_{i+1} \dots x_n},
\end{equation*}
which corresponds to the definition of the QFT with reversed order of the qubits.

The \emph{inverse QFT} is depicted in Figure~\ref{fig: QFT}~(b). The different gate groups are the inverse operators of $\G_1 ,\dots, \G_n$ as given above. 
Therefore, each gate group $\G^{-1}_i$ of the inverse QFT consists of the adjoints, i.e., the conjugate transposes, of the controlled phase-shift gates in $\G_i$ and a Hadamard gate. 
Using the core notation, we can express $\G_i^{-1}$ as
\begin{equation*}
  \G^{-1}_i = \frac{1}{\sqrt{2}} I^{\otimes(i-1)}  \otimes \core{ \begin{bmatrix}1 & 0 \\ 1 & 0 \end{bmatrix} & \begin{bmatrix}0 & 1 \\ 0 & -1 \end{bmatrix}} \otimes \core{I &  \\  & R^*_2 } \otimes \dots \otimes \core{I &  \\  & R^*_{n-i}} \otimes \core{I \\ R^*_{n-i+1}}.
\end{equation*}
Analogously to the QFT, we can show that $\G_i^{-1} \ket{\psi}$  has a rank-one representation if $\psi_j$ is either $0$ or $1$ for $j=1, \dots, n$. We then have
\begin{equation*}
 \mathrm{QFT}^{-1} \ket{\psi} = \frac{1}{\sqrt{N}} \sum_{y=0}^{N-1} e^{-\frac{2 \pi \mathrm{i}}{N} \psi y} \ket{y}= \frac{1}{\sqrt{N}} \left(\ket{0} + e^{-2 \pi \mathrm{i} 0.\psi_n } \ket{1} \right) \otimes \dots \otimes \left(\ket{0} + e^{-2 \pi \mathrm{i} 0.\psi_1 \dots \psi_n } \ket{1} \right).
\end{equation*}
In Section~\ref{sec: Shor's algorithm}, we will employ the inverse QFT in the context of Shor's algorithm.

\section{Numerical experiments}
\label{sec: Numerical experiments}

In this section, we will provide numerical illustrations of the theoretical results presented in this work. 
The algorithms have been implemented in Python 3.8 and added to the toolbox Scikit-TT\footnote{\url{https://github.com/PGelss/scikit_tt}}.
For comparison purposes, we use Qiskit\footnote{\url{https://github.com/Qiskit/qiskit}}~\cite{Qiskit2021} for simulating quantum circuits.

\subsection{Quantum full adder network}
\label{sec: Quantum full adder network}

\begin{figure}[htbp]
  \centering
  \begin{subfigure}[b]{0.47\textwidth}
   \centering
    \begin{quantikz}[row sep={0.6cm,between origins}]
        \lstick{$\ket{C_\textrm{in}}$} & \gate[4]{\text{QFA}} & \qw                  & \qw & \qw & \rstick{$\ket{S_1}$}\qw \\
        \lstick{$\ket{A_1}$}         & \qw                  & \qw                  & \qw & \qw & \rstick{$\ket{A_1}$}\qw \\
        \lstick{$\ket{B_1}$}         & \qw                  & \qw                  & \qw & \qw & \rstick{$\ket{B_1}$}\qw \\
        \lstick{$\ket{0}$}           & \qw                  & \gate[4]{\text{QFA}} & \qw & \qw & \rstick{$\ket{S_2}$}\qw \\
        \lstick{$\ket{A_2}$}         & \qw                  & \qw                  & \qw & \qw & \rstick{$\ket{A_2}$}\qw \\
        \lstick{$\ket{B_2}$}         & \qw                  & \qw                  & \qw & \qw & \rstick{$\ket{B_2}$}\qw \\
        \lstick{$\ket{0}$}           & \qw                  & \qw                  & \qw & & \\[-0.2cm]
        &                      &                      & \vdots & & \\
        &                      &     &     & \gate[4]{\text{QFA}} & \rstick{$\ket{S_n}$}\qw \\
        \lstick{$\ket{A_n}$}         & \qw                  & \qw & \qw &               & \rstick{$\ket{A_n}$}\qw \\
        \lstick{$\ket{B_n}$}         & \qw                  & \qw & \qw &               & \rstick{$\ket{B_n}$}\qw \\
        \lstick{$\ket{0}$}           & \qw                  & \qw & \qw &               & \rstick{$\ket{C_\textrm{out}}$}\qw 
    \end{quantikz}
    \caption{}
  \end{subfigure}%
  \hfill
  \begin{subfigure}[b]{0.47\textwidth}
   \centering
     \begin{quantikz}[row sep={0.6cm,between origins}]
        \lstick{$\ket{0}$}      & \qw & \gate[12, nwires={7,8,9}]{\text{QFAN}} & \meter{} \\
        \lstick{$\ket{0}$}              & \gate{H} & \qw                    & \qw \\
        \lstick{$\ket{0}$}              & \gate{H} & \qw                    & \qw \\
        \lstick{$\ket{0}$}                & \qw & \qw                    & \meter{} \\
        \lstick{$\ket{0}$}              & \gate{H} & \qw       & \qw \\
        \lstick{$\ket{0}$}              & \gate{H} & \qw                    & \qw \\
        \lstick{$\ket{0}$}                & \qw & \qw                    & \rstick{\raisebox{-0.6cm}{$\vdots$}}\\[-0.2cm]
        \lstick{\raisebox{-1cm}{$\vdots$}}&     &                        & \\
                                          &     &                        & \meter{} \\
        \lstick{$\ket{0}$}              & \gate{H} & \qw       & \qw \\
        \lstick{$\ket{0}$}              & \gate{H} & \qw                    & \qw \\
        \lstick{$\ket{0}$}                & \qw & \qw                    & \meter{}
    \end{quantikz}
    \caption{}
  \end{subfigure}
  \caption{Quantum full-adder network: (a) A QFAN is constructed by successively coupling several QFAs, i.e., by passing the last output qubit of one adder to the first input qubit of another adder. (b)~The QFANs simulated in this experiment act on qubit registers of different sizes with a superposition of the inputs for $A_i$, $B_i$ as initial quantum state.}
  \label{fig: QFA network}
\end{figure}
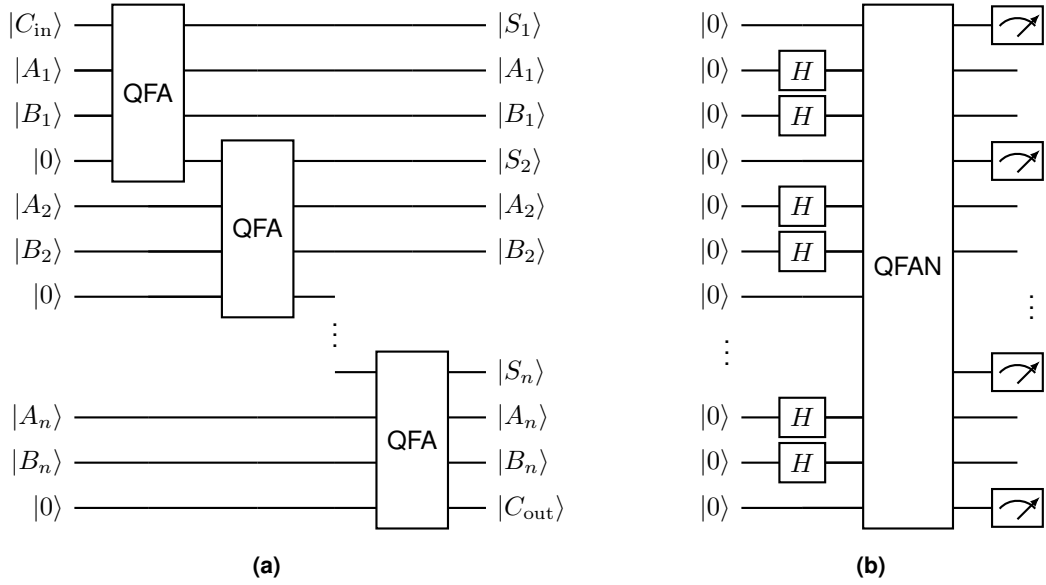

As a first example, we consider a full-adder network. In Section~\ref{sec: Quantum full adder} we already described how to express a single quantum full adder as an MPO. Due to the structure of the decomposition, we can directly couple several adders in order to construct a quantum full-adder network (QFAN). More precisely, the corresponding MPO representation of the network can be derived by contracting the last core of the previous adder with the first core of the subsequent adder:

\begin{equation*}
\begin{split}
 \G_\textrm{QFAN} &= \prod_{i=1}^n I^{\otimes 3(n-i)} \otimes \G \otimes I^{\otimes 3(i-1)}\\
 &= \left( I^{\otimes 3(n -1)} \otimes \G \right) \cdot \left( I^{\otimes 3(n -2)} \otimes \G \otimes I^{\otimes 3} \right) \cdot  \ldots  \cdot \left(\G \otimes I^{\otimes 3(n-1)} \right)\\
 & = \core{ \G^{(1)} } \otimes \core{ \G^{(2)} } \otimes \core{ \G^{(3)} } \otimes \core{\mathbf{C}} \otimes  \core{ \G^{(2)} } \otimes \core{ \G^{(3)} } \otimes \core{\mathbf{C}} \otimes \dots  \\
 &\qquad \ldots \otimes \core{\mathbf{C}} \otimes \core{ \G^{(2)} } \otimes \core{ \G^{(3)} } \otimes \core{ \G^{(4)} } ,
\end{split}
\end{equation*}
where $\G$ denotes the MPO~\eqref{eq: QFA MPO} and $n$ is the number of QFAs in the network, i.e., the size of the qubit system is $3n+1$. The core $\mathbf{C}$ is given by
\begin{equation*}
 \core{\mathbf{C}} = \core{ \G^{(1)}_\textrm{QFA} } \cdot \core{ \G^{(4)}_\textrm{QFA} } = \core{\sigma_x C_0 & I & \sigma_x C_1 } \cdot \core{I \\ \sigma_x} = \core{\sigma_x C_0 & I & \sigma_x C_1 \\ C_1 & \sigma_x & C_0}.
\end{equation*}
Here, $\cdot$ denotes the (core-wise) contraction of corresponding column and row modes. The diagrammatic notation of the quantum full-adder network is shown in Figure~\ref{fig: QFA network}~(a).

\begin{figure}[htbp]
  \centering
  \begin{subfigure}[b]{0.475\textwidth}
   \centering
    \includegraphics[width=7cm]{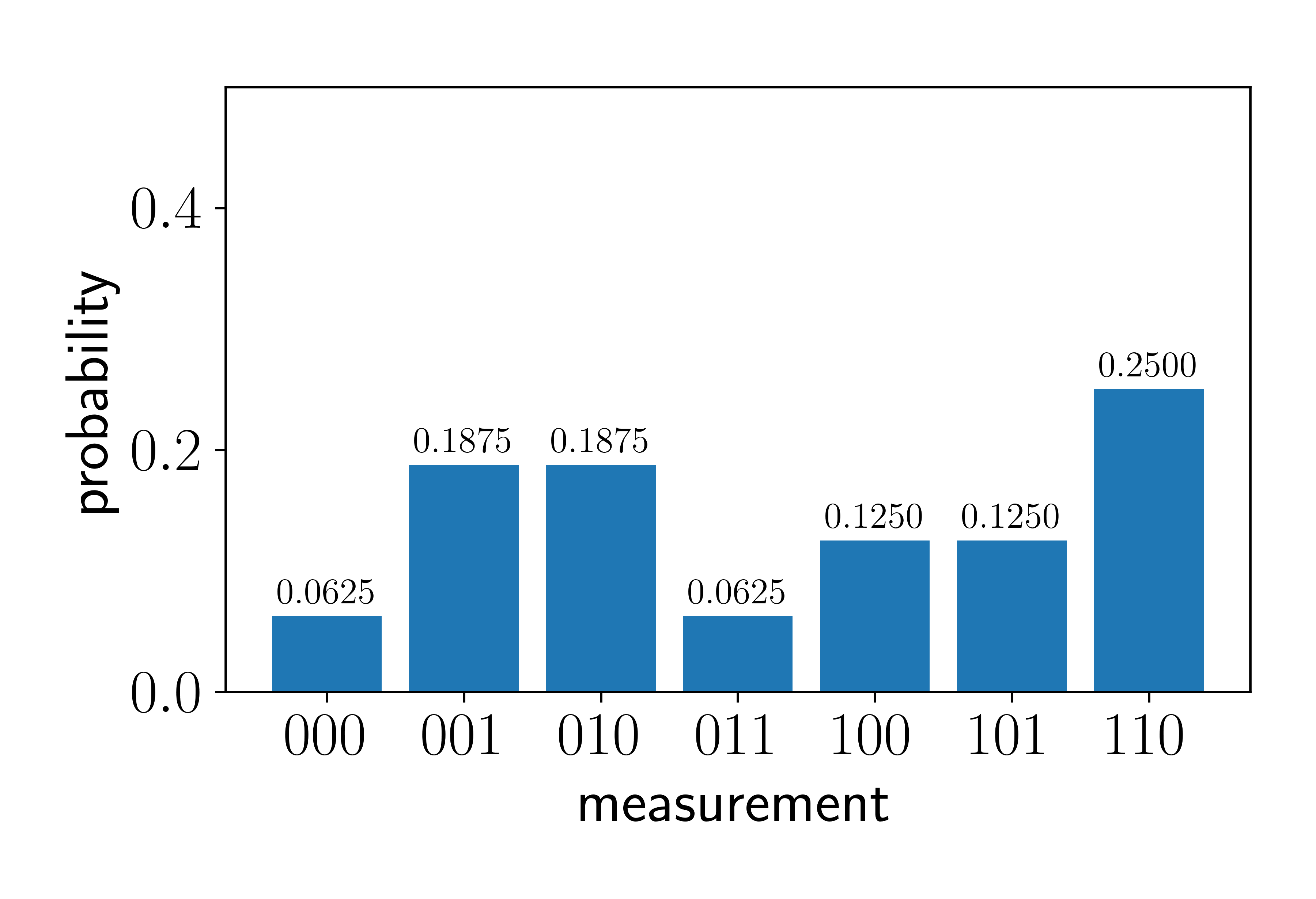}
    \caption{}
  \end{subfigure}
  \hfill
  \begin{subfigure}[b]{0.475\textwidth}
   \centering
     \includegraphics[width=7cm]{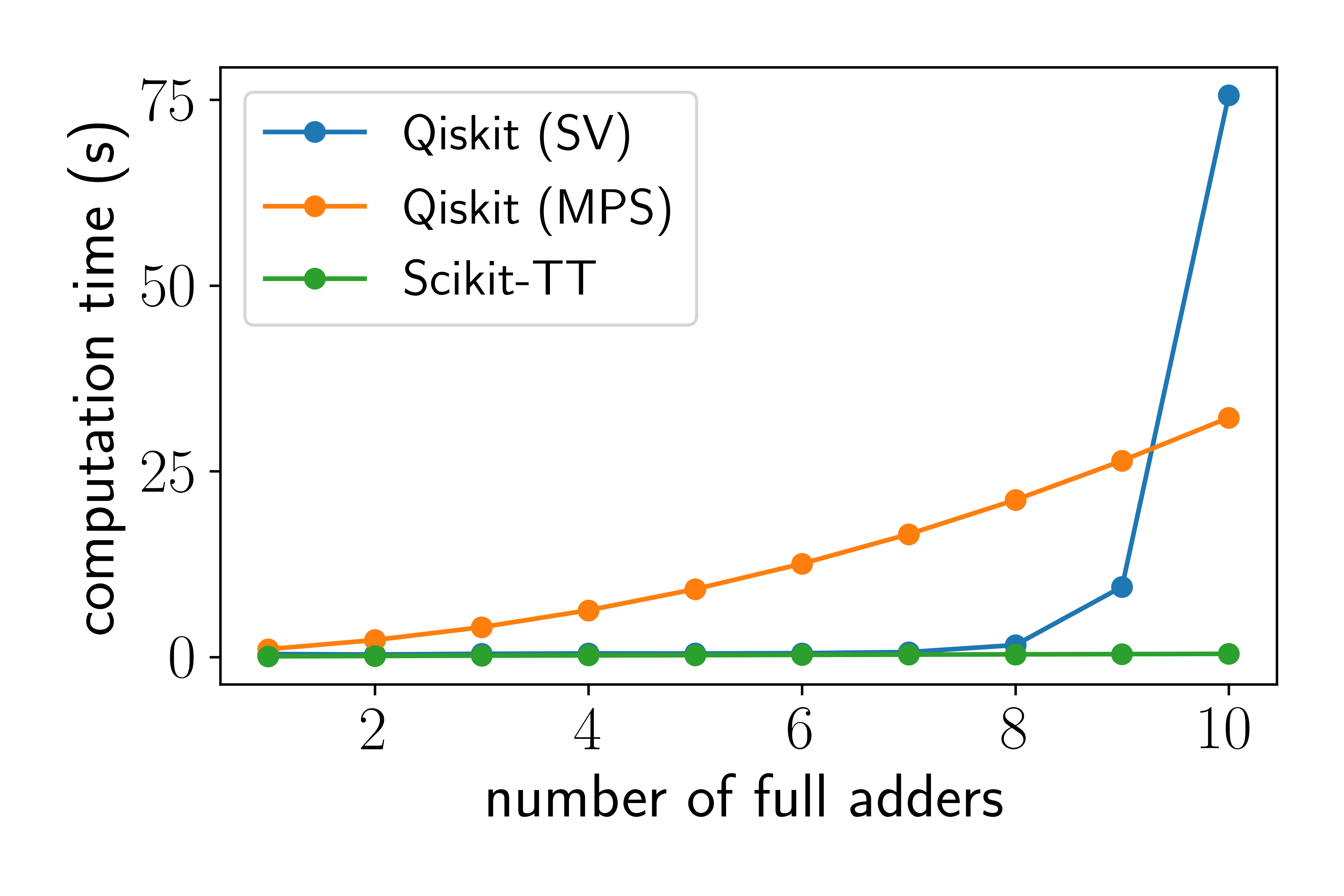}
    \caption{}
  \end{subfigure}
  \caption{Results of the QFAN simulations: (a) Example of the probability distribution when measuring $\ket{S_1, \dots, S_n, C_\textrm{out}}$ for $n=2$, computed by generative sampling of the final quantum state in MPS format. (b) Average computation times for generating $10^5$ samples for different numbers of QFAs using Qiskit (either with the statevector (SV) or MPS simulation method) and Scikit-TT.}
  \label{fig: QFAN results}
\end{figure}

We implement quantum full-adder networks comprising different numbers of full adders and compare our results with the state probabilities computed by Qiskit when measuring the output qubits $S_i$, \mbox{$i=1, \dots, n$}, and $C_\textrm{out}$. 
In order to not only add simple basis states but rather superposition states, we choose the following state as the initial state:
\begin{equation*}
\begin{split}
 \PSI &= (I \otimes H \otimes H \otimes I \otimes H \otimes H \otimes I \otimes \dots \otimes I \otimes H \otimes H \otimes I) \cdot \ket{0^{3n+1}} \\
 &= \frac{1}{2^n} \begin{bmatrix} 1 \\ 0 \end{bmatrix} \otimes \begin{bmatrix} 1 \\ 1 \end{bmatrix} \otimes \begin{bmatrix} 1 \\ 1 \end{bmatrix} \otimes \begin{bmatrix} 1 \\ 0 \end{bmatrix} \otimes \begin{bmatrix} 1 \\ 1 \end{bmatrix} \otimes \begin{bmatrix} 1 \\ 1 \end{bmatrix} \otimes \begin{bmatrix} 1 \\ 0 \end{bmatrix} \otimes \dots \otimes \begin{bmatrix} 1 \\ 0 \end{bmatrix} \otimes \begin{bmatrix} 1 \\ 1 \end{bmatrix} \otimes \begin{bmatrix} 1 \\ 1 \end{bmatrix} \otimes \begin{bmatrix} 1 \\ 0 \end{bmatrix},
\end{split}
\end{equation*}
where Hadamard gates are applied to each of the input qubits $\ket{A_i}$ and $\ket{B_i}$, $i=1, \dots, n$, cf.~Figure~\ref{fig: QFA network}~(b). 
For manageable sizes of the QFAN, we are able to directly extract the probability tensor $\PROB$ from the product $\G_\mathrm{QFAN} \PSI$, see Section~\ref{sec: Probability distributions and generative sampling}. 
Our experiments confirm that the distributions computed by Qiskit converge to those given by $\PROB$ (disregarding different orderings) with increasing sample size. See Figure~\ref{fig: QFAN results}~(a) for the probability distribution for $n = 2$.
However, since the storage consumption of the probability tensor is $O(2^{n+1})$ in the full format, we employ the sampling strategy described in Section~\ref{sec: Probability distributions and generative sampling}. 
In Figure~\ref{fig: QFAN results}~(b), we show the average computation times (including construction and measuring phase) needed for generating $10^5$ samples of the output of QFANs with increasing number of full adders.

For the Qiskit simulations, we use the AerSimulator backend and employ both the statevector as well as the MPS simulation method.
That is, similar to our method, Qiskit also uses MPS networks for expressing the quantum states, whereas gate operations on non-consecutive qubits require a series of swap gates applied to the MPS cores to ensure that the respective qubits are adjacent.
One can see that, in contrast to the Qiskit statevector simulations, the runtimes of the tensor-based approaches depend only linearly on the number of QFAs.
The MPO-based method implemented using Scikit-TT, however, requires significantly less time for constructing and sampling the final quantum state.
This example demonstrates the potential of tensor formats for speeding up the sampling phase in quantum system simulations.
Moreover, the proposed MPO-based approach may help to mitigate the curse of dimensionality especially when considering circuits on high-dimensional qubit registers.
For instance, we can easily simulate a network consisting of $100$ QFAs and only need about $30$\,s for generating $10^6$ samples.

\subsection{Quantum Fourier transform}

As illustrated in the previous section, one of the possible advantages of MPO-based circuit representations is the acceleration of quantum simulations in certain cases. 
In order to support this claim, let us consider the MPO decomposition of the QFT introduced in Section~\ref{sec: Quantum Fourier transform}.
Here, we rely on the decompositions of the gate groups and do not consider a closed MPO representation of the whole circuit at once.
In fact, the operators $\G_i$ corresponding to the different gate groups all have an MPO rank of $2$ (except for $\G_n$ which has rank $1$) and their contraction would yield a highly-entangled MPO.
That is, we simulate the QFT by alternatingly applying an operator $\G_i$~\eqref{eq: QFT gate group} and performing an orthonormalization step, see Appendix~\ref{app: orthonormalization}, to control the ranks of the intermediate quantum states.
As we already discussed in Section~\ref{sec: Quantum Fourier transform}, every gate gate group $\G_i$ maps to a quantum state which can be represented by a rank-one tensor if the quantum register is initialized to a basis state in the computational basis.
For the following experiment, we therefore randomly choose initial quantum states from the computational basis while increasing the size of the qubit system and compare Scikit-TT with Qiskit simulations using the built-in MPS simulation method.
The results of the different quantum circuit simulations in terms of computation times for generating $s = 10^2, 10^4, 10^6$ samples are shown in Table~\ref{tab: QFT}.

\begin{table}[htbp]
\centering
\renewcommand{\arraystretch}{1.15}
\setlength{\tabcolsep}{0.5cm}
\caption{Results of QFT simulations: For different numbers of qubits $n$ and samples $s$, the average execution times (in seconds) and corresponding standard deviations of the Qiskit and Scikit-TT simulations are shown. For each parameter choice, the simulations were repeated $100$ times with randomly drawn basis states as inputs.}
\begin{tabular}{llrrrr}
\hline
 ~         &           & $n=16$ & $n=32$ & $n=64$ & $n=128$ \\ \hline 
 \multirow{3}{*}{\rotatebox[origin=c]{90}{Qiskit}} & \multicolumn{1}{|l}{$s=10^2$}    & $0.07 \pm 0.08$    & $0.12 \pm 0.02$    & $0.33 \pm 0.05$    & $1.09 \pm 0.11$\\
 & \multicolumn{1}{|l}{$s=10^4$}    & $0.16 \pm 0.01$    & $0.27 \pm 0.03$    & $0.57 \pm 0.05$    & $1.57 \pm 0.10$\\
 & \multicolumn{1}{|l}{$s=10^6$}    & $6.65 \pm 0.03$    & $15.12 \pm 0.05$    & $25.72 \pm 0.10$    & $48.35 \pm 0.11$\\\hline
 \multirow{3}{*}{\rotatebox[origin=c]{90}{Scikit-TT}} & \multicolumn{1}{|l}{$s=10^2$} & $0.04 \pm 0.00$    & $0.15 \pm 0.00$    & $0.59 \pm 0.00$    & $2.31 \pm 0.01$ \\
           & \multicolumn{1}{|l}{$s=10^4$} & $0.07 \pm 0.00$    & $0.20 \pm 0.00$    & $0.66 \pm 0.00$    & $2.43 \pm 0.01$ \\
           & \multicolumn{1}{|l}{$s=10^6$} & $5.34 \pm 0.15$    & $7.82 \pm 0.18$    & $11.51 \pm 0.16$    & $20.39 \pm 0.20$ \\\hline
\end{tabular}
\label{tab: QFT}
\end{table}

\noindent While the computation time of Qiskit grows (almost) linearly with the size of the qubit system, the scaling behavior of the MPO approach changes with increasing number of samples.
This can be explained by the low-rank MPS/MPO representations of the quantum states and gate groups:
In addition to the sampling procedure, the dominant operations in our simulations are the orthonormalization procedures, i.e., applying sequences of $n$ SVDs to every MPS decomposition of the $n$ intermediate states, see Appendix~\ref{app: orthonormalization}.
Since the ranks of every gate group operator and, therefore, every intermediate quantum state are bounded by $2$, the computational complexity of the construction of the final quantum state can be estimated as $O(8 n^2)$.
For the sampling procedure, we can estimate the computational costs as $O(8 s n)$, cf.~Section~\ref{sec: Probability distributions and generative sampling}.
That is, the complexity is mainly determined by the construction and orthonormalization of the wave functions in MPS format for a small number of samples and, thus, increases quadratically with the number of qubits.
On the other hand, the computation times show a linearly scaling with the system size when we consider 
larger numbers of samples because the computational complexity of the sampling scheme outweighs that of the orthonormalization steps.
For high-dimensional qubit systems, in particular, a large number of required measurements is a common bottleneck of many quantum algorithms.
Thus, MPO/MPS-based formulations not only offer a alternative way of treating quantum circuits but may also accelerate the simulation of these if the final quantum state has sufficiently low MPS ranks.

\subsection{Shor's algorithm}
\label{sec: Shor's algorithm}

Shor's algorithm~\cite{Shor1994} was one of the first quantum algorithms that provided a significant advantage over classical computations.
Given an integer $M$, the goal is to find a non-trivial divisor by combining classical and quantum computations. 
The computational complexity of Shor's algorithm is $O((\log M)^3)$, demonstrating an almost exponential acceleration compared to best-known classical factorization algorithms.
Therefore, Shor’s algorithms might be used to break RSA encryption~\cite{Gidney2021}---which is based on composite numbers having two large prime factors---much faster than in the classical case.
The classical part of Shor’s algorithm is to find a number $a$ that has no common divisor with a given composite number $M$, whereas the quantum part exploits the inverse QFT on a $2n$-qubit register with $N = 2^n > M$, see Section~\ref{sec: Quantum Fourier transform}, for finding the period $p$ of the function $f(x) = a^x \bmod M$ in polynomial time.
The structure of Shor's algorithm can be summarized by the following essential stages:
\begin{enumerate}
 \item[1)] Choose $1 < a < M$ randomly. If $a$ and $M$ are coprime, go to step 2, otherwise $\mathrm{gcd}(a,M)$ is a nontrivial factor of $M$.
 \item[2)] Initialize the input register with $2n$ qubits and the target register with $n$ qubits, both to the zero state. Apply Hadamard gates to the qubits in the input register, the modular exponentiation circuit $U_f$ to both registers and afterwards the inverse QFT to the input register. When measuring the input register, we obtain a (basis) state $\ket{y}$.
 \item[3)] Calculate candidates for the period $p$ from $y$, see below. If all candidates fail, go back to step 2.
 \item[4)] If $p$ is odd or $a^{p/2} \equiv -1\bmod {N} $, got back to step 1.  Otherwise, either $\mathrm{gcd}(a^{p/2}-1, M)$ or $\mathrm{gcd}(a^{p/2}+1, M)$ is a nontrivial factor of $M$.
\end{enumerate}

\begin{figure}[htbp]
  \centering
     \begin{quantikz}[row sep={0.6cm,between origins}]
        &\lstick{$\ket{0}_1$} & \gate{H} & \gate[6, nwires={2,5}]{~~U_f~~} & \gate[3, nwires={2}]{~~\mathrm{QFT}^{-1}~~}  &  \meter{} \\
        &\lstick{$\vdots$~~~}    & \vdots   &   &    & \vdots \\[0.1cm]
        &\lstick{$\ket{0}_1$} & \gate{H} &           & \qw  &  \meter{} \\
        &\lstick{$\ket{0}_2$} & \qw & & \qw & \qw\\
        &\lstick{$\vdots$~~~}                &    &    &  &  \\[0.1cm]
        &\lstick{$\ket{0}_2$} &  \qw & & \qw & \qw \\
    \end{quantikz}
  \caption{Shor's algorithm: The input register (marked by the subscript $1$) holds the superposition of the basis states $\ket{0}, \dots, \ket{N^2-1}$, while the output register (marked the by subscript $2$) holds the superposition of the values of $f$ after application of $U_f$. The inverse QFT is applied to the first register to estimate the phase of $U_f$.}
  \label{fig: Shor}
\end{figure}
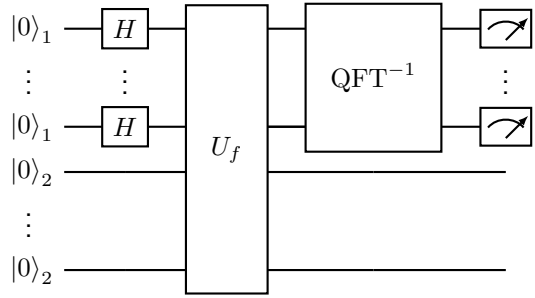

\noindent Figure~\ref{fig: Shor} shows the corresponding quantum circuit.
The first step, computing the greatest common divisor of $a$ and $M$, can be simply performed by using the Euclidean algorithm.
The modular exponentiation circuit $U_f$ in the second step represents the function $f$ in the sense that the quantum state after the application of the Hadamard gates and $U_f$ is 
\begin{equation}\label{eq: modular exponentiation}
 U_f \left(  \frac{1}{N} \sum_{x=0}^{N^2-1} \ket{x, 0^n }\right) = \frac{1}{N} \sum_{x=0}^{N-1} \ket{x, f(x)}.
\end{equation}
Applying the inverse QFT to the first register leads to the quantum state
\begin{equation*}
 \mathrm{QFT}^{-1} \left( \frac{1}{N} \sum_{x=0}^{N^2-1} \ket{x, f(x)} \right) = \frac{1}{N^2} \sum_{x=0}^{N^2-1} \sum_{y=0}^{N^2 -1} e^{-\frac{2 \pi \mathrm{i}}{N^2} x y}\ket{y, f(x)}.
\end{equation*}
Supposing we finally obtain a state $\ket{y}$ when measuring the first register, we can use continued fraction expansion on a classical computer to find approximations $z/q$ of $y/N$, where $q$ is equal to the period $p$ or at least a factor of it with high probability. For a more detailed description of the different steps of Shor's algorithm, we refer to~\cite{Nielsen2010}. 
The construction of $U_f$ is the most challenging part when implementing Shor's algorithm. 
It is usually based on controlled modular multiplier blocks, individually designed for each choice of $a$ and $M$~\cite{Pavlidis2014, Duan2020}.
In terms of tensor products, however, we can construct an explicit representation for any base and modulus.
That is, with respect to the above circuit, the tensor operator corresponding to~\eqref{eq: modular exponentiation} is given by 
\begin{equation*}
\begin{split}
 \mathbf{U}_f &= \sum_{x=0}^{N^2-1} C_{x_1} \otimes \dots \otimes C_{x_{2n}} \otimes \sigma_x^{f(x)_1} \otimes \dots \otimes \sigma_x^{f(x)_n} \\
 &= \sum_{z=0}^{N-1} \left( \sum_{\substack{x =1\\ f(x)=z}}^{N^2-1}C_{x_1} \otimes \dots \otimes C_{x_{2n}} \right) \otimes \sigma_x^{z_1} \otimes \dots \otimes \sigma_x^{z_n},
\end{split}
\end{equation*}
where we again define $C_0 = I - C$ and $C_1 = C$. 

Let us consider the case of $M=15$ and $a \in \{2,4,7,8,11,13\}$. 
That is, our input register has $8$ and the target register $4$ qubits.
In fact, the operator $\mathbf{U}_f$ corresponding to any choice of $a$ can be represented as an MPO with either rank $2$ or $4$, which is a direct consequence of the multiplicative orders of the different bases.
The explicit decompositions are given by
\begin{equation*}
\begin{array}{ l @{} c @{} c @{} c @{} c @{} c @{} c @{} c @{} c @{} c @{} c @{} c @{} c @{} c @{} c @{} c @{} c}
a=2:  & \qquad \mathbf{U}_f  & ~ = ~& ~ & I^{\otimes 6} & ~\otimes~ & C_0 & ~ \otimes ~ & C_0 & ~\otimes~ & I & ~\otimes~ & I & ~\otimes~ & I & ~\otimes~ & \sigma_x\\
      &                    &   & +~~ & I^{\otimes 6} & ~\otimes~ & C_0 & ~ \otimes ~ & C_1 & ~\otimes~ & I & ~\otimes~ & I & ~\otimes~ & \sigma_x & ~\otimes~ & I\\
      &                    &   & +~~ & I^{\otimes 6} & ~\otimes~ & C_1 & ~ \otimes ~ & C_0 & ~\otimes~ & I & ~\otimes~ & \sigma_x & ~\otimes~ & I & ~\otimes~ & I\\
      &                    &   & +~~ & I^{\otimes 6} & ~\otimes~ & C_1 & ~ \otimes ~ & C_1 & ~\otimes~ & \sigma_x & ~\otimes~ & I & ~\otimes~ & I & ~\otimes~ & I,\\[0.25cm]
a=4:  & \qquad \mathbf{U}_f  & ~ = ~& ~ & I^{\otimes 6} & ~\otimes~ & I & ~ \otimes ~ & C_0 & ~\otimes~ & I & ~\otimes~ & I & ~\otimes~ & I & ~\otimes~ & \sigma_x\\
      &                    &   & +~~ & I^{\otimes 6} & ~\otimes~ & I & ~ \otimes ~ & C_1 & ~\otimes~ & I & ~\otimes~ & \sigma_x & ~\otimes~ & I & ~\otimes~ & I,\\[0.25cm]
a=7:  & \qquad \mathbf{U}_f  & ~ = ~& ~ & I^{\otimes 6} & ~\otimes~ & C_0 & ~ \otimes ~ & C_0 & ~\otimes~ & I & ~\otimes~ & I & ~\otimes~ & I & ~\otimes~ & \sigma_x\\
      &                    &   & +~~ & I^{\otimes 6} & ~\otimes~ & C_0 & ~ \otimes ~ & C_1 & ~\otimes~ & I & ~\otimes~ & \sigma_x & ~\otimes~ & \sigma_x & ~\otimes~ & \sigma_x\\
      &                    &   & +~~ & I^{\otimes 6} & ~\otimes~ & C_1 & ~ \otimes ~ & C_0 & ~\otimes~ & I & ~\otimes~ & \sigma_x & ~\otimes~ & I & ~\otimes~ & I\\
      &                    &   & +~~ & I^{\otimes 6} & ~\otimes~ & C_1 & ~ \otimes ~ & C_1 & ~\otimes~ & \sigma_x & ~\otimes~ & \sigma_x & ~\otimes~ & I & ~\otimes~ & \sigma_x,\\[0.25cm]
a=8:  & \qquad \mathbf{U}_f  & ~ = ~& ~ & I^{\otimes 6} & ~\otimes~ & C_0 & ~ \otimes ~ & C_0 & ~\otimes~ & I & ~\otimes~ & I & ~\otimes~ & I & ~\otimes~ & \sigma_x\\
      &                    &   & +~~ & I^{\otimes 6} & ~\otimes~ & C_0 & ~ \otimes ~ & C_1 & ~\otimes~ & \sigma_x & ~\otimes~ & I & ~\otimes~ & I & ~\otimes~ & I\\
      &                    &   & +~~ & I^{\otimes 6} & ~\otimes~ & C_1 & ~ \otimes ~ & C_0 & ~\otimes~ & I & ~\otimes~ & \sigma_x & ~\otimes~ & I & ~\otimes~ & I\\
      &                    &   & +~~ & I^{\otimes 6} & ~\otimes~ & C_1 & ~ \otimes ~ & C_1 & ~\otimes~ & I & ~\otimes~ & I & ~\otimes~ & \sigma_x & ~\otimes~ & I,\\[0.25cm]
a=11:  & \qquad \mathbf{U}_f  & ~ = ~& ~ & I^{\otimes 6} & ~\otimes~ & I & ~ \otimes ~ & C_0 & ~\otimes~ & I & ~\otimes~ & I & ~\otimes~ & I & ~\otimes~ & \sigma_x\\
      &                    &   & +~~ & I^{\otimes 6} & ~\otimes~ & I & ~ \otimes ~ & C_1 & ~\otimes~ & \sigma_x & ~\otimes~ & I & ~\otimes~ & \sigma_x & ~\otimes~ & \sigma_x,\\[0.25cm]
a=13:  & \qquad \mathbf{U}_f  & ~ = ~& ~ & I^{\otimes 6} & ~\otimes~ & C_0 & ~ \otimes ~ & C_0 & ~\otimes~ & I & ~\otimes~ & I & ~\otimes~ & I & ~\otimes~ & \sigma_x\\
      &                    &   & +~~ & I^{\otimes 6} & ~\otimes~ & C_0 & ~ \otimes ~ & C_1 & ~\otimes~ & \sigma_x & ~\otimes~ & \sigma_x & ~\otimes~ & I & ~\otimes~ & \sigma_x\\
      &                    &   & +~~ & I^{\otimes 6} & ~\otimes~ & C_1 & ~ \otimes ~ & C_0 & ~\otimes~ & I & ~\otimes~ & \sigma_x & ~\otimes~ & I & ~\otimes~ & I\\
      &                    &   & +~~ & I^{\otimes 6} & ~\otimes~ & C_1 & ~ \otimes ~ & C_1 & ~\otimes~ & I & ~\otimes~ & \sigma_x & ~\otimes~ & \sigma_x & ~\otimes~ & \sigma_x,\\[0.25cm]
a=14:  & \qquad \mathbf{U}_f  & ~ = ~& ~ & I^{\otimes 6} & ~\otimes~ & I & ~ \otimes ~ & C_0 & ~\otimes~ & I & ~\otimes~ & I & ~\otimes~ & I & ~\otimes~ & \sigma_x\\
      &                    &   & +~~ & I^{\otimes 6} & ~\otimes~ & I & ~ \otimes ~ & C_1 & ~\otimes~ & \sigma_x & ~\otimes~ & \sigma_x & ~\otimes~ & \sigma_x & ~\otimes~ & I.
\end{array}
\end{equation*}

Thus, we can represent Shor's algorithm by a sequence of MPOs comprising a superpositioning operator, the operator $\mathbf{U}_f$, and the gate groups of the inverse QFT, see Section~\ref{sec: Quantum Fourier transform}.
Note that the quantum state after the application the first two MPOs is given by
\begin{equation*}
 \mathbf{U}_f  \left( H^{\otimes 2n} \otimes I^{\otimes n} \right) \ket{0^{2n},0^n} = \frac{1}{2^n} \mathbf{U}_f \cdot \left( \begin{bmatrix}1 \\ 1 \end{bmatrix} \otimes \dots \otimes \begin{bmatrix}1 \\ 1 \end{bmatrix} \otimes \begin{bmatrix}1 \\ 0 \end{bmatrix} \otimes \dots \otimes \begin{bmatrix}1 \\ 0 \end{bmatrix}
\right)
\end{equation*}
and, therefore, has the same ranks (at most) as $\mathbf{U}_f$. 
As in the previous section, we orthonormalize the intermediate quantum states to reduce the ranks between the applications of the gate groups of the inverse QFT, see Appendix~\ref{app: orthonormalization}.
Lastly, we construct the probability tensor as described in Section~\ref{sec: Probability distributions and generative sampling}.
Reading the measured bitstrings in reversed order (due to the omission of SWAP operations), we then find for each choice of $a$ that the results shown in Table~\ref{tab: Shor} are exactly the same as when using the Qiskit implementation described in~\cite{Abbas2020}.

\begin{table}[htbp]
\centering
\renewcommand{\arraystretch}{1.15}
\setlength{\tabcolsep}{0.5cm}
\caption{Results of Shor's factorization algorithm for $M=15$: The possible bases can be divided in two groups. For $a \in \{2,7,8,13\}$, the final quantum state (before measuring) has rank $r=4$ and the measured basis state indices remain the same for each $a$. If $a \in \{4,11,14\}$, then the final quantum state can be represented by a rank-$2$ MPS. For both groups, the extracted period and the corresponding prime factors are shown. If $y=0$ is measured, Shor's algorithm fails and returns no factors of $M$.}
\begin{tabular}{ccccc}
\hline
base $a$ & rank $r$ & measurement $y$ & period $q$ & factors $(M_1, M_2)$  \\ \hline 
\multirow{ 4}{*}{$2,7,8,13$}& \multirow{ 4}{*}{$4$} & $0$ & $1$  &  $\varnothing$ \\
& &  $64$ & $4$  &  $(3,5)$ \\
& &  $128$ & $2$  &  $(3,1)$ \\
& &  $192$ & $4$  &  $(3,5)$ \\\hline
\multirow{ 2}{*}{$4,11,14$}& \multirow{ 2}{*}{$2$} &  $0$& $1$  & $\varnothing$  \\
& &   $128$& $2$  & $(3,1)$  \\\hline
\end{tabular}
\label{tab: Shor}
\end{table}

The results show that it is possible to construct low-rank decompositions for even more complex quantum algorithms. Although we here do not focus on an explicit representation of the entire circuit, we can directly simulate its action on the qubit register while keeping the ranks of the intermediate quantum states low. The ranks of the final quantum state in this case reflect the order of the associated base.

\section{Conclusion and outlook}
\label{sec: Conclusion and outlook}

Simulating quantum circuits on classical computers is crucial for designing new quantum algorithms, but quickly leads to huge storage consumption and computational costs.
The development of more efficient simulation techniques requires the exploitation of suitable mathematical concepts.

This study presents an approach that represents quantum circuits as MPOs and can be regarded as a generalization of the description of quantum states as MPSs.
We have shown with the aid of various benchmark problems that MPOs are a systematic way to express (networks of) quantum logic gates acting on arbitrary qubits in a closed form.
We showed that it is possible to derive compact MPO representations of various quantum circuits with ranks much smaller than the theoretical bounds, allowing for easily accessible interpretations in terms of entanglement and their action on different basis states.
Our numerical experiments showed that the low-rank structure of MPO decompositions can be exploited to speed up computations while yielding equivalent results as in classical simulations of quantum circuits.
So far, we only investigated ideal quantum simulations, i.e., concepts like error correction or fault tolerance were not considered yet.
However, the focus of our future research will shift from the realm of ideal quantum computers to noisy ones.

The presented results are an essential step towards the MPO-based construction and simulation of quantum circuits.
As MPSs are a natural way of describing quantum states, we showed that MPOs can be used in the same fashion for representing quantum circuits in a closed form and demonstrated a new framework for designing and simulating complex quantum circuits.
Therefore, future studies will particularly focus on the relationships between MPO decompositions and quantum circuit representations using sets of universal gates.

\section*{Acknowledgments}

This research has been funded by the Deutsche Forschungsgemeinschaft (CRC 1114, \emph{``Scaling Cascades in Complex Systems''}).

\printbibliography
\appendix

\section{Appendix}
\label{app: Appendix}

\subsection{Construction and manipulation of MPO cores}

\subsubsection{Linear transformations of MPS cores}
\label{app: core manipulation}

Given an MPS core $\T^{(i)} \in \C^{r_{i-1} \times d_i \times r_i}$ as defined in~\eqref{eq: core notation - single core}, applying a linear transformation represented by the matrix $Q \in \C^{r_i \times \tilde{r}_i}$ from the right results in a new core $\U^{(i)} \in \C^{r_{i-1} \times d_i \times \tilde{r}_i}$, i.e.,
\begin{equation}\label{eq: core manipulation}
 \left\llbracket \U^{(i)} \right\rrbracket = \left\llbracket \T^{(i)} \right\rrbracket \cdot Q = \left\llbracket \T^{(i)} \right\rrbracket \cdot \begin{bmatrix} Q_{1,1} & \cdots & Q_{1,\tilde{r}_i} \\
 \vdots & \ddots & \vdots \\  Q_{r_i,1} & \cdots & Q_{r_i,\tilde{r}_i}  \end{bmatrix},
\end{equation}
where each core entry of $\U^{(i)}$ is given by 
\begin{equation*}
 \U^{(i)}_{k,x,\ell} = \sum_{\mu=1}^{r_i} \T^{(i)}_{k,x,\mu} Q_{\mu,\ell}.
\end{equation*}
That is, the notation in~\eqref{eq: core manipulation} is an alternative way of expressing TT cores defined by the relation
\begin{equation*}
 \U^{(i)}_{:,x,:} = \T^{(i)}_{:,x,:} Q
\end{equation*}
for $x \in \{1, \dots, d_i\}$. The left-multiplication of a matrix $Q \in \C^{\tilde{r}_{i-1} \times r_{i-1}}$ and a TT core and the core manipulation of MPOs 
are defined analogously. 
For two TT cores $\T^{(i)} \in \C^{r_{i-1} \times d_i \times r_i}$, $\T^{(i+1)} \in \C^{\tilde{r}_{i} \times d_{i+1} \times r_{i+1}}$ and a matrix $Q \in \C^{r_i \times \tilde{r}_i}$, it holds that
\begin{equation}\label{eq: core manipulation 2}
\left( \left\llbracket \T^{(i)} \right\rrbracket \cdot Q \right) \otimes \left\llbracket \T^{(i+1)} \right\rrbracket = \left\llbracket \T^{(i)} \right\rrbracket \otimes \left( Q \cdot \left\llbracket \T^{(i+1)} \right\rrbracket \right).
\end{equation}
This can be directly shown by considering the representation of a single entry of an MPS, see~\eqref{eq: MPS - single entry}.

\subsubsection{Orthonormalization of tensors in MPS format}
\label{app: orthonormalization}

The left- and right-unfoldings of an MPS core $\T^{(i)}$ are given by the matrices $\mathcal{L}\in \C^{(r_{i-1} \cdot d_i) \times r_i}$ and $\mathcal{R} \in \C^{r_{i-1} \times ( d_i \cdot r_i)}$ with $\mathcal{L}_{\overline{k,x},\ell} = \mathcal{R}_{k, \overline{x, \ell}} = \T^{(i)}_{k,x,\ell}$. Here, the indices of two modes of $\T^{(i)}$ are lumped into the multi-index $\overline{k,x}$ and $\overline{x, \ell}$, respectively, typically based on reverse lexicographic or colexicographic ordering. The remaining mode forms the other dimension of the unfolding matrix. We call the core $\T^{(i)}$ left-orthonormal if its left-unfolding is orthonormal with respect to the rows, i.e., $\mathcal{L}^\top  \mathcal{L} = I \in \R^{r_i \times r_i}$. Correspondingly, a core is called right-orthonormal if its right-unfolding is orthonormal with respect to the columns, i.e., \mbox{$\mathcal{R} \mathcal{R}^\top = I \in \R^{r_{i-1} \times r_{i-1}}$}. In graphical notation, orthonormal components are depicted by half-filled circles, see Figure~\ref{fig: probability tensors}\,(c).

In order to orthonormalize a given MPS with ranks $(r_0 , \dots , r_n)$, we can apply a sequence of (truncated) SVDs to the cores of the decomposition.
A right-orthonormal MPS representation, for example, can be obtained by iteratively computing the SVD $U \Sigma V^\top$ of the $i$th core and contracting the non-orthonormal part (i.e., $U \Sigma$) with the $(i-1)$th core which is then decomposed in the next step. 
The (reshaped) matrix $V^\top$ builds the updated core $\tilde{\T}^{(i)}$.
Starting at the last MPS core $\T^{(n)}$ and repeating this procedure until reaching $\T^{(2)}$ results in a right-orthonormalized MPS decomposition $\tilde{\T}$ with cores $\tilde{\T}^{(i)} \in \C^{s_{i-1} \times d_i \times s_i}$ as required for the sampling technique described in Section~\ref{sec: Probability distributions and generative sampling}. 
In fact, it holds that $\tilde{r}_i \leq r_i$ for $i =0 , \dots, n$, i.e., the ranks may become smaller while orthonormalizing if certain left- and right-unfoldings, respectively, exhibit a low-rank structure.
Furthermore, truncated SVDs can be used to reduce the ranks by using absolute or relative cut-off criteria for the singular values, cf.~\cite{Oseledets2009b}.
In general, the computational complexity of left- and right-orthonormalizations, which are also known in the context of \emph{Schmidt decompositions} of quantum systems, can be estimated as $O(n \, d \, r^3)$, where $d$ denotes the maximum mode dimension and $r$ the maximum rank of of $\T$.

\subsubsection{Diagonal tensors as MPOs}
\label{sec: Diagonal tensors as MPOs}

Given a tensor $\T \in \C^{d_1 \times \dots \times d_n}$, we define the corresponding diagonal tensor $\diag(\T) \in \C^{D \times D}$ by 
\begin{equation*}
 \diag(\T)_{x_1, \dots, x_n, y_1, \dots, y_n} = \begin{cases} \T_{x_1, \dots , x_n}, & \text{if}~x_i = y_i~\text{for}~i=1,\dots,n, \\ 0, & \text{otherwise.} \end{cases}
\end{equation*}
If $\T$ is given as an MPS with cores $\T^{(i)} \in \C^{r_{i-1} \times d_i \times r_i}$, then $\diag(\T)$ can be written as an MPO with cores defined by
\begin{equation*}
 \diag(\T)^{(i)}_{k_{i-1}, x_i,y_i k_i } = \begin{cases} \T^{(i)}_{k_{i-1}, x_i, k_i}, & \text{if}~x_i = y_i, \\ 0, & \text{otherwise.} \end{cases}
\end{equation*}
Just as in standard linear algebra, it holds that $\diag(\T) \cdot \T = \T \odot \T$, where $\odot$ denotes the (multidimensional) Hadamard product.

\subsection{MPO representations of quantum circuits}

\subsubsection{Quantum full adder}
\label{app: QFA}

In order to derive the MPO representation~\eqref{eq: QFA MPO}, we consider the product~\eqref{eq: QFA product} and concatenate the quantum gates successively. 
The product $\G_1 = \mathrm{CNOT}(2 \mid 3) \cdot \mathrm{CCNOT}(2,3 \mid 4)$ is given by
\begin{equation*}
\begin{split}
 \G_1 &=  \left( I  \otimes  \core{ I & C }  \otimes  \core{ I \\ \sigma_x - I  }  \otimes  I \right) \cdot \left(  I  \otimes  \core{ I & C}  \otimes  \core{ I & 0 \\ 0 & C  }  \otimes  \core{ I \\ \sigma_x -I } \right)\\
 &= I  \otimes  \core{ I & C & C & C^2 }  \otimes  \core{ I & 0  \\ 0 & C \\ \sigma_x - I & 0 \\ 0 & (\sigma_x - I)C }  \otimes  \core{ I \\ \sigma_x - I }.
\end{split}
\end{equation*}
Since $C^2 = C$, we can add the last three rows of the third core and get
\begin{equation*}
\begin{split}
 \G_1 &= I  \otimes  \core{ I & C }  \otimes  \core{ I & 0  \\ \sigma_x - I & \sigma_x C  } \otimes  \core{ I \\ \sigma_x - I }.
\end{split}
\end{equation*}
For the concatenation of the $\mathrm{CNOT}(3 \mid 1)$ and $\mathrm{CCNOT}(1, 3 \mid 4)$ gates, we have
\begin{equation*}
\begin{split}
 \G_2 & = \mathrm{CNOT}(3 \mid 1) \cdot \mathrm{CCNOT}(1,3 \mid 4)\\ &=  \left( \core{ I & \sigma_x - I }  \otimes  \core{ I & 0 \\ 0 & I }  \otimes  \core{ I \\ C  }  \otimes  I \right) \cdot \left( \core{ I & C }  \otimes  \core{ I & 0 \\ 0 & I }  \otimes  \core{ I & 0 \\ 0 & C  }  \otimes  \core{ I \\ \sigma_x -I } \right)\\
 &= \core{ I & C & \sigma_x - I & (\sigma_x-I)C }  \otimes  \core{ I & 0 & 0 & 0 \\ 0 & I & 0 & 0 \\ 0 & 0 & I & 0 \\ 0 & 0 & 0 & I }  \otimes  \core{ I & 0  \\ 0 & C \\ C & 0 \\ 0 & C }  \otimes  \core{ I \\ \sigma_x - I }\\
 &= \core{ I & C & \sigma_x - I & (\sigma_x-I)C }  \otimes  \core{ I & 0 & 0  \\ 0 & I & 0  \\ 0 & 0 & I  \\ 0 & I & 0  }  \otimes  \core{ I & 0  \\ 0 & C \\ C & 0  }  \otimes  \core{ I \\ \sigma_x - I }\\
 &= \core{ I & \sigma_x C & \sigma_x - I }  \otimes  \core{ I & 0 & 0  \\ 0 & I & 0  \\ 0 & 0 & I  }  \otimes  \core{ I & 0  \\ 0 & C \\ C & 0  }  \otimes  \core{ I \\ \sigma_x - I }.
\end{split}
\end{equation*}
In order to bring the first core into the desired form, cf.~\eqref{eq: QFA MPO}, we linearly combine the columns by multiplying with a (non-singular) matrix from the right. Therefore, we have to multiply the second core with the inverse of that matrix from the left to preserve the given MPO, i.e.,
\begin{gather*}
 \left(\core{ I & \sigma_x C & \sigma_x - I } \cdot \begin{bmatrix} 1 & 1 & 0 \\ -1 & 0 & 1 \\ 1 & 0 & 0 \end{bmatrix} \right) \otimes \left(  \begin{bmatrix} 0 & 0 & 1 \\ 1 & 0 & -1 \\ 0 & 1 & 1 \end{bmatrix} \cdot \core{ I & 0 & 0  \\ 0 & I & 0  \\ 0 & 0 & I  }  \right)\\
 = \core{ \sigma_x C_0 & I & \sigma_x C_1 } \otimes \core{ 0 & 0 & I  \\ I & 0 & -I  \\ 0 & I & I  },
\end{gather*}
where we use $C_0 = I-C$ and $C_1 = C$.
Combining the above MPOs, we obtain 
\begin{equation*}
\begin{split}
 \G_2 \cdot \G_1 &= \core{ \sigma_x C_0 & I & \sigma_x C_1 }  \otimes  \core{ 0 & 0 & 0 & 0 & I & C_1  \\ I & C_1 & 0 & 0 & -I & -C_1  \\ 0 & 0 & I & C_1 & I & C_1  } \\
 & \qquad \otimes  \core{ I & 0 & 0 & 0  \\ \sigma_x-I & \sigma_x C_1 & 0 & 0 \\ 0 & 0 & C_1 & 0 \\ 0 & 0 & C_1 (\sigma_x -I ) & C_1 \sigma_x C_1 \\ C_1 & 0 & 0 & 0\\ C_1 (\sigma_x -I ) & C_1 \sigma_x C_1 & 0 & 0 }  \otimes  \core{ I \\ \sigma_x - I \\ \sigma_x - I \\ (\sigma_x - I)^2 }.
\end{split}
\end{equation*}
Since $C_1 \sigma_x C_1 = 0$, it follows that
\begin{equation*}
\begin{split}
 \G_2 \cdot \G_1 
 &= \core{ \sigma_x C_0 & I & \sigma_x C_1 }  \otimes  \core{ 0 & 0 & 0 & 0 & I & C_1  \\ I & C_1 & 0 & 0 & -I & -C_1  \\ 0 & 0 & I & C_1 & I & C_1  } \\
 & \qquad \otimes  \core{ I & 0    \\ \sigma_x-I & \sigma_x C_1   \\ 0 & C_1  \\ 0 & C_1 (\sigma_x -I )  \\ C_1 & 0  \\ C_1 (\sigma_x -I ) & 0   }  \otimes  \core{ I \\ \sigma_x - I  }.
\end{split}
\end{equation*}
Now, we manipulate the second core by
\begin{equation*}
\begin{split}
 \core{ 0 & 0 & 0 & 0 & I & C_1  \\ I & C_1 & 0 & 0 & -I & -C_1  \\ 0 & 0 & I & C_1 & I & C_1  } \cdot \begin{bmatrix} 1 & 0 & 1 & 0 & 0 & 0 \\ -1 & 1 & -1 & 1 & 0 & 0 \\ -1 & 0 & 0 & 0 & 1 & 0 \\ 1 & -1 &  0 & 0 & -1 & 1  \\ 1 & 0 & 0 &  0 & 0 & 0 \\ -1 & 1 & 0 & 0 &  0 & 0 \end{bmatrix} = \core{ C_0 & C_1 & 0 & 0 & 0 & 0  \\ 0 & 0 & C_0 & C_1 & 0 & 0  \\ 0 & 0 & 0 & 0 & C_0 & C_1  },
\end{split}
\end{equation*}
and the last core by 
\begin{equation*}
\begin{split}
  \begin{bmatrix}1 & 0 \\ 1 & 1 \end{bmatrix} \cdot \core{I \\ \sigma_x - I} = \core{ I \\ \sigma_x }.
\end{split}
\end{equation*}
Thus, the third core becomes
\begin{gather*}
 \begin{bmatrix} 0 & 0 & 0 & 0 & 1 & 0 \\ 0 & 0 & 0 & 0 & 1 & 1 \\ 1 & 0 & 0 & 0 & -1 & 0 \\ 1 & 1 &  0 & 0 & -1 & -1  \\ 0 & 0 & 1 & 0 & 1 & 0 \\ 0 & 0 & 1 & 1 &  1 & 1 \end{bmatrix} \cdot \core{ I & 0    \\ \sigma_x-I & \sigma_x C_1   \\ 0 & C_1  \\ 0 & C_1 (\sigma_x -I )  \\ C_1 & 0  \\ C_1 (\sigma_x -I ) & 0   } \cdot \begin{bmatrix} 1 & 0 \\ -1 & 1\end{bmatrix} = \core{ C_1 & 0    \\ C_1 \sigma_x & 0   \\ C_0 & 0  \\ 0  & \sigma_x C_1  \\ 0 & C_1  \\ 0 & C_1 \sigma_x  }.
\end{gather*}
Thus, we can express the MPO operator $\G$~\eqref{eq: QFA product} as the product
\begin{equation*}
\begin{split}
 \G &= \mathrm{CNOT}(2 \mid 3) \cdot \G_2 \cdot \G_1 = \left( I  \otimes  \core{ C_0 & C_1} \otimes \core{ I \\ \sigma_x  } \otimes  I \right) \cdot \G_2 \cdot \G_1.
\end{split}
\end{equation*}
Since the first and last core of $\mathrm{CNOT} (2 \mid 3)$ are simply given by the identity matrix, let us only consider the second and third core of the product. We obtain
\begin{equation*}
\begin{split}
  \left\llbracket \G^{(2)} \right\rrbracket \otimes \left\llbracket \G^{(3)} \right\rrbracket&= \core{ 
  C_0 & 0 & 0   & 0 & 0   & 0 & 0 & C_1 & 0 & 0   & 0 & 0 \\ 
  0   & 0 & C_0 & 0 & 0   & 0 & 0 & 0   & 0 & C_1 & 0 & 0 \\ 
  0   & 0 & 0   & 0 & C_0 & 0 & 0 & 0   & 0 & 0   & 0 & C_1  } \otimes 
  \core{ C_1 & 0    \\ C_1 \sigma_x & 0   \\ C_0 & 0  \\ 0  & \sigma_x C_1  \\ 0 & C_1  \\ 0 & C_1 \sigma_x \\ \sigma_x C_1 & 0    \\ C_0 & 0   \\ \sigma_x C_0 & 0  \\ 0  & C_1  \\ 0 & \sigma_x C_1  \\ 0 & C_0 } \\
  &= \core{ 
  C_0  & C_1  & 0   & 0 \\ 
  0    & C_0  & C_1 & 0 \\ 
  0    & 0    & C_0 & C_1  } \otimes 
  \core{ C_1 & 0   \\ C_0 & 0  \\ 0 & C_1  \\ 0 & C_0 },
\end{split}
\end{equation*}
which leads to the decomposition~\eqref{eq: QFA MPO}.

\subsubsection{Simon's algorithm}
\label{app: Simon}

We have 
\begin{equation*}
 \G_1 = \G_4 = H \otimes I \otimes H \otimes I \otimes H \otimes I \otimes H \otimes I.
\end{equation*}
Furthermore, it holds that
\begin{equation*}
\begin{array}{c@{}c@{}c@{}c@{}c@{}c@{}c@{}c@{}c@{}c@{}c@{}c@{}c@{}c@{}c@{}c@{}c}
\G_3 \, = & ~& \core{ I & C} & ~\otimes~ & \core{ I & \\ & I } & ~\otimes~ & \core{ I & \\ & I } & ~ \otimes ~ & \core{ I & \\ & I } & ~\otimes~ & \core{ I & \\ & I} & ~\otimes~ & \core{ I \\ \sigma_x - I } & ~\otimes~ & I & ~ \otimes ~ & I\\[0.3cm]
& \hspace*{0.25cm}\cdot\hspace*{0.25cm} & \core{ I & C} & ~\otimes~ & \core{ I \\ \sigma_x - I } & ~\otimes~ & I & ~ \otimes ~ & I & ~\otimes~ &  I & ~\otimes~ & I & ~\otimes~ & I & ~ \otimes ~ & I,
\end{array}
\end{equation*}
which can be rewritten as
\begin{equation*}
\begin{array}{r@{}c@{}c@{}c@{}c@{}c@{}c@{}c@{}c@{}c@{}c@{}c@{}c@{}c@{}c@{}c@{}c}
\G_3 \, = & ~& \core{ C_0 & C_1} & ~\otimes~ & \core{ I & \\ & I } & ~\otimes~ & \core{ I & \\ & I } & ~ \otimes ~ & \core{ I & \\ & I } & ~\otimes~ & \core{ I & \\ & I} & ~\otimes~ & \core{ I \\ \sigma_x } & ~\otimes~ & I & ~ \otimes ~ & I\\[0.3cm]
& \hspace*{0.25cm}\cdot\hspace*{0.25cm} & \core{ C_0 & C_1} & ~\otimes~ & \core{ I \\ \sigma_x } & ~\otimes~ & I & ~ \otimes ~ & I & ~\otimes~ &  I & ~\otimes~ & I & ~\otimes~ & I & ~ \otimes ~ & I\\[0.3cm]
=& ~ & \core{ C_0 & C_1} & ~\otimes~ & \core{ I & \\ & \sigma_x } & ~\otimes~ & \core{ I & \\ & I } & ~ \otimes ~ & \core{ I & \\ & I } & ~\otimes~ &  \core{ I & \\ & I } & ~\otimes~ & \core{ I \\ \sigma_x} & ~\otimes~ & I & ~ \otimes ~ & I,
\end{array}
\end{equation*}
see Example~\ref{ex: W state and CNOT gate as TT}. The product of $\G_3$ and $\G_2$ (see Section~\ref{sec: Simons algorithm}) can then be expressed as
\begin{equation*}
\begin{split}
 \G_3 \cdot \G_2 &= \core{C_0 & C_1} \otimes \core{I & \\ & I} \otimes \core{C_0 & C_1 & & \\ & & C_0 & C_1} \otimes \core{I & \\ \sigma_x & \\ & I \\ & \sigma_x} \\
 &\qquad \otimes \core{C_0 & C_1 \\ C_1 & C_0} \otimes \core{I \\ \sigma_x} \otimes \core{C_0 & C_1} \otimes \core{I \\ \sigma_x}.
\end{split}
\end{equation*}
Multiplying the cores with odd indices by $H$ from left and right results in the MPO~\eqref{eq: Simon - gates}.

\subsubsection{Quantum Fourier transform}
\label{app: QFT}

As already mentioned in Section~\ref{sec: Quantum gates}, we can swap the roles of the control and target qubit for a controlled phase-shift gate. That is, by repeatedly applying~\eqref{eq: core manipulation 2}, we have
\begin{equation*}
\begin{split}
 \mathrm{CPHASE}_k(p \mid q) &= I^{\otimes (p-1)} \otimes \core{C_0 & C_1} \otimes \core{I^{\otimes (q-p-1)} & \\ & I^{\otimes (q-p-1)}} \otimes \core{I \\ R_k} \otimes I^{\otimes (n-q)}\\
 &= I^{\otimes (p-1)} \otimes \core{C_0 & C_1} \otimes \core{I^{\otimes (q-p-1)} & \\ & I^{\otimes (q-p-1)}} \\
 & \qquad \otimes \left( \begin{bmatrix} 1 & 1 \\ 1 & e^{\frac{2 \pi \mathrm{i}}{2^k}} \end{bmatrix} \cdot \core{C_0 \\ C_1} \right)\otimes I^{\otimes (n-q)}\\
 &= I^{\otimes (p-1)} \otimes \left( \core{C_0 & C_1} \cdot \begin{bmatrix} 1 & 1 \\ 1 & e^{\frac{2 \pi \mathrm{i}}{2^k}} \end{bmatrix} \right) \otimes \core{I^{\otimes (q-p-1)} & \\ & I^{\otimes (q-p-1)}} \\
 & \qquad \otimes \core{C_0 \\ C_1} \otimes I^{\otimes (n-q)}\\
 &= I^{\otimes (p-1)} \otimes \core{I & R_k} \otimes \core{I^{\otimes (q-p-1)} & \\ & I^{\otimes (q-p-1)}} \otimes \core{C_0 \\ C_1} \otimes I^{\otimes (n-q)}\\
 &= \mathrm{CPHASE}_k(q \mid p),
\end{split}
\end{equation*}
where $\mathrm{CPHASE}_k (p \mid q)$ denotes a controlled phase-shift gate as described in Section~\ref{sec: Quantum Fourier transform}, which applies $R_k$ to the target qubit $q$ if the control qubit $p$ is in state $\ket{1}$. 
Thus, each gate group $\G_i$, $i=1, \dots, n-1$, of the QFT can be written as
\begin{equation*}
\begin{split}
 \G_i &=  \mathrm{CPHASE_{n-i+1}(n,i)} \cdot \ldots \cdot \mathrm{CPHASE_2(i+1,i)} \cdot \left( I^{\otimes(i-1)} \otimes H \otimes I^{\otimes (n-i)}\right)\\
 &=  \mathrm{CPHASE_{n-i+1}(i,n)} \cdot \ldots \cdot \mathrm{CPHASE_2(i,i+1)} \cdot \left( I^{\otimes(i-1)} \otimes H \otimes I^{\otimes (n-i)}\right)\\
 &= \left( I^{\otimes(i-1)} \otimes \core{C_0 & C_1} \otimes \core{I & \\ & R_2} \otimes \dots \otimes \core{I \\ R_{n-i+1}} \right) \cdot  \left( I^{\otimes(i-1)} \otimes H \otimes I^{\otimes (n-i)}\right) \\
 &=  \frac{1}{\sqrt{2}} I^{\otimes(i-1)} \otimes \core{ \begin{bmatrix} 1 & 1 \\ 0 & 0\end{bmatrix} & \begin{bmatrix} 0 & 0 \\ 1 & -1\end{bmatrix}} \otimes \core{I & \\ & R_2} \otimes \dots \otimes \core{I & \\ & R_{n-i}} \otimes \core{I \\ R_{n-i+1}}.
\end{split}
\end{equation*}
The MPO representation of the inverse QFT can be derived analogously.

\end{document}